\documentclass[aps,prd,twocolumn,superscriptaddress,nofootinbib,amsmath,amssymb,floatfix]{revtex4-2}

\usepackage{graphicx}
\usepackage{bm}
\usepackage{dcolumn}
\usepackage{multirow}
\usepackage[colorlinks=true,linkcolor=blue,citecolor=blue,urlcolor=black]{hyperref}

\makeatletter
\renewcommand{\p@subsection}{\thesection\,}
\makeatother

\newcommand{\nB}{n_B}
\newcommand{\ce}{c_e}
\newcommand{\cs}{c_s}
\newcommand{\mut}{\tilde\mu}
\newcommand{\Msun}{M_\odot}
\newcommand{\fm}{\,\mathrm{fm}^{-3}}
\newcommand{\mev}{\,\mathrm{MeV}}
\newcommand{\km}{\,\mathrm{km}}
\newcommand{\bv}{Brunt--V\"ais\"al\"a}
\newcommand{\taugw}{\tau_{\rm GW}}
\newcommand{\Lbar}{\bar\Lambda}
\newcommand{\fdyn}{f_{\rm dyn}}
\newcommand{\Sst}{S_{\rm struc}}
\newcommand{\Scomp}{S_{\rm comp}}

\begin{document}

\title{Quarkyonic matter suppresses neutron-star $g$ modes
       and reverses their mass trend}

\author{Probit J Kalita}
\affiliation{Department of Physics and Astronomy,
National Institute of Technology, Rourkela 769008, India}

\author{Bharat Kumar}
\email{kumarbh@nitrkl.ac.in}
\affiliation{Department of Physics and Astronomy,
National Institute of Technology, Rourkela 769008, India}

\date{\today}

\begin{abstract}
Gravity ($g$) modes are the only neutron-star oscillations that report on the
composition of dense matter rather than on its stiffness, and work on hybrid
stars has established that a first-order quark transition raises their
frequency. Quarkyonic matter does the opposite. Quark--nucleon chemical
equilibrium is maintained by the strong interaction, so the quarks acquire no
thermodynamic freedom on an oscillation period and $\cs^2-\ce^2$ reduces to a
positive-definite quadratic form in the lepton gradients alone. The nucleon
momentum shell stiffens both sound speeds together instead of separating them,
so the buoyancy factor collapses by a factor of $9.5$ to $19$ at an early
transition and the core is left only weakly stratified. Solving the $l=2$
relativistic Cowling problem for ten equations of state that share one
isoscalar sector, seven quarkyonic and three nucleonic controls at matched
symmetry-energy slope, we found the $g$ mode confined to the nucleonic shell
outside the core, with the horizontal flow that it lives on excluded from the
core while the core is displaced almost rigidly. Its frequency falls from
$158$--$522$~Hz across the controls to $81$--$254$~Hz across the quarkyonic
models, and by $14\%$ at $1.4\,\Msun$ for the matched pair at $L=50\mev$; more
important than the shift, it \emph{decreases} with mass where the controls
rise. Writing
each frequency as the dynamical frequency $(GM/R^3)^{1/2}$ times a
dimensionless remainder separates structure from composition. The $f$ and $p_1$
trends prove to be inherited from the mass--radius relation; the $g_1$
remainder, constant to $2$--$9\%$ along a nucleonic sequence, falls by a
quarter to a third along a quarkyonic one and departs by a factor of two from
the nucleonic $g$-mode relation at fixed compactness and $L/K_0$. These are
Cowling values, lower bounds at the ten-per-cent level, and the sign of the
trend survives a correction of that size.
\end{abstract}

\maketitle

\section{Introduction}
\label{s:intro}

The bulk properties of a neutron star are now measured well enough to
constrain the pressure of cold dense matter. Radio timing has established
pulsars of two solar masses~\cite{Antoniadis13,Fonseca21}, NICER has delivered
radii for several sources~\cite{Salmi24,Choudhury24,Miller21}, and the tidal
deformability inferred from GW170817 bounds the pressure around twice
saturation density~\cite{Abbott18}. What none of these measures is what the
pressure is made of. A mass--radius relation is an integral of
$p(\varepsilon)$, so interiors of quite different composition can be arranged
to lie on the same curve, and the construction studied here is a case in point:
quarkyonic matter reproduces ordinary stellar structure while replacing much of
the interior with deconfined quarks~\cite{Zhao20}. Oscillation spectra offer a
way past the degeneracy, because a mode frequency depends on how the matter is
layered as well as on how stiff it is~\cite{Finn87,Kokkotas01}, and one family
of modes depends on the layering alone. This paper asks what that family does
when the layering is produced by a quarkyonic momentum shell.

Such a family exists because dense matter has two sound speeds rather than
one. Along the weak-equilibrium sequence the slope of
$p(\varepsilon)$ is the \emph{equilibrium} sound speed $\ce$. A fluid element
displaced on a millisecond cannot re-equilibrate everything; it re-equilibrates
only what is fast compared with the mode period and responds instead with the
\emph{adiabatic} sound speed $\cs$, evaluated at frozen slow variables. The
buoyancy that supports gravity ($g$) modes is carried entirely by the
difference, through the \bv\ frequency
$N^2=g^2(\ce^{-2}-\cs^{-2})e^{\nu-\lambda}$, and in ordinary $npe\mu$ matter
the slow variable is the lepton content~\cite{Reisenegger92,Lai94}. Close the
gap between the two speeds and the restoring force vanishes with it.
Quarkyonic matter, we found, very nearly closes it: the fundamental $g$-mode
frequency fell from $158$--$522$~Hz in our nucleonic controls to
$81$--$254$~Hz, and its trend with mass reversed.

That a cold star rings because its composition changes with depth is an old
idea~\cite{Finn87,Reisenegger92}, and its application to quark matter is not
new~\cite{Fu08,Wei20}. What the modern proposals share is not a lineage but a
mechanism, and that mechanism has one requirement: a discontinuity. A
first-order deconfinement transition, or a Gibbs mixed phase, drops $\ce$
abruptly while leaving $\cs$ untouched, so that $\ce^{-2}-\cs^{-2}$ spikes and
$f_{g_1}$ moves from the $0.2$--$0.4$~kHz of a purely hadronic star to
$0.6$--$0.9$~kHz~\cite{Zheng23}. A sharp interface does the same more
violently, because no mixed region smooths the
jump~\cite{Jaikumar21,Constantinou21}. Hyperons~\cite{Tran23} and a
dark-matter component~\cite{Shirke25} put a different constituent through the
same argument. In all of these the signal is large because something falls
discontinuously, and a construction with nothing discontinuous anywhere in it
lies outside the scope of every one of them.

How far such a signal can be trusted is itself disputed, and our results bear
on the disagreement. Sotani and Togashi~\cite{SotaniTogashi26} fit the
nucleonic $f_{g_1}$ to compactness and to $\eta_0=L/K_0$ at the ten-per-cent
level, and propose a departure from that fit as the signature of new
composition. Kuan \emph{et al.}~\cite{Kuan22}, working in full general
relativity with a larger library of equations of state, find that the $g$-mode
space does not collapse onto one relation at all, but splits into groups
labelled by the class of equation of state. What is at issue is whether there
is a $g$-mode universal relation to be violated in the first place.
Section~\ref{s:univ} takes a side, and Eq.~(\ref{e:whyuniv}) says why.

Quarkyonic matter poses the composition question and answers it without any
discontinuity at all. The name was coined by McLerran and
Pisarski~\cite{McLerran07}, who argued that at large $N_c$ there is a regime
that is confined, in the sense that its excitations are baryons, and yet
quark-like, in the sense that its thermodynamics is that of a quark Fermi sea.
In the neutron-star realisation of McLerran and Reddy~\cite{McLerran19} that
compromise becomes concrete: nucleons are expelled from the interior of their
own Fermi sphere and confined to a shell $k_{0i}\le k\le k_{Fi}$, while quarks
fill the space they vacate. Forced to sit at high momentum, the nucleons carry
more kinetic energy than they otherwise would and the pressure climbs steeply: a rapid but \emph{continuous} stiffening, a peak in $\ce^2$ and not a drop.
That is what reconciles a soft equation of state (EOS) near saturation with the
existence of two-solar-mass stars, and it is why the construction is of
interest. Zhao and Lattimer~\cite{Zhao20} placed it on a thermodynamically
consistent footing by imposing genuine chemical equilibrium between quarks and
nucleons and solving for $\beta$-equilibrated $npdue\mu$ matter; theirs is the
model we adopt.

What a $g$ mode does in such matter has not been computed, and it cannot be
computed by feeding a new EOS into an existing pipeline, because the
\emph{definition} of $\cs$ changes. The reason is a separation of rates that
can be checked before any code is written. Quarkyonic matter carries two
chemical equilibria. The quark--nucleon conditions
$\mu_d=(2\mu_n-\mu_p)/3$ and $\mu_u=(2\mu_p-\mu_n)/3$ are enforced by the
strong interaction, on a timescale $\hbar/\Lambda_{\rm QCD}\sim10^{-23}$~s; the
lepton fractions are set by modified-Urca processes, on timescales of seconds
to hours. The $g$-mode periods computed here run from $1.9$ to $12.4$~ms. The
strong conditions are therefore faster than the mode by twenty-one orders of
magnitude and the weak ones slower by at least three, so both limits are clean
and neither needs a finite-rate treatment: the adiabatic derivative must be
taken at fixed lepton fractions with strong equilibrium re-imposed at every
step. Neither the standard $npe\mu$ prescription (freeze the proton fraction) nor
the mixed-phase prescription of Refs.~\cite{Jaikumar21,Zheng23} (freeze the
particle fractions within each phase and let the volume fraction relax) is the
correct limit here. Oscillations of quarkyonic stars have been
computed once before, by Dey \emph{et al.}~\cite{Dey26}, but for spacetime
$\omega$ modes, which are indifferent to composition and never require the two
sound speeds to be told apart.

A single equation of state cannot settle a question of this kind, so we worked
with a family. All ten models share one isoscalar sector and one saturation
point, $K_0=220\mev$ and $S_v=31\mev$; what varies across the family is the
symmetry-energy slope $L$, the shell-transparency scale $\Lambda$ and the
transition density $n_t$. Seven are quarkyonic and three are the purely
nucleonic Zhao--Lattimer functionals~\cite{Zhao20} with the same $L$ values, so
that every quarkyonic star can be read against a control that differs from it
in the momentum shell and in nothing else (Table~\ref{t:eos}); all ten are
spliced to the same SLy crust~\cite{Read09}. Any difference that
survives that pairing is the shell, and any spread that survives the family is
theoretical uncertainty rather than a result.

The physical answer can be stated before the machinery is assembled. A
quarkyonic core is only weakly stratified, for two reasons that act together. The
quark sea is chemically slaved to the nucleons, so that when a fluid element is
compressed the quarks absorb most of the chemical imbalance that would
otherwise restore it; and the shell structure stiffens the matter so rapidly
that the imbalance which does survive generates very little buoyancy. Both
speeds rise together instead of separating, and the buoyancy factor
$\ce^{-2}-\cs^{-2}$ fell by a factor of $9.5$ to $19$ across the early-onset
models. The $g$ mode is then expelled from the core into the nucleonic shell
around it, and the heavier the star, the thinner that shell. The suppressed
frequency, the reversed mass trend, the lengthened damping time and the small
strain all follow from that.

Separating this from structure needs care, because a quarkyonic star differs
from a nucleonic one in composition and in structure at once. We therefore
wrote each frequency as the star's dynamical frequency $(GM/R^3)^{1/2}$ times a
dimensionless remainder. The split is an identity rather than a model, and it
is the instrument the rest of the paper uses: what survives in the remainder is
composition, because the structure has been divided out. Figure~\ref{f:univ},
which plots that remainder against compactness for every model in the family,
is the figure the argument rests on. Measured this way, the $f$ and $p_1$ modes
of a quarkyonic star are unremarkable: their remainders vary with mass much
as a nucleonic star's do, and their anomalous mass trends are inherited from
the mass--radius relation, which is itself anomalous, since these stars
\emph{expand} as they gain mass where the controls contract. The $g_1$
remainder does not behave that way. It is constant to $2$--$9\%$ along a
nucleonic sequence, so that a hadronic $g$ mode is little more than the
dynamical frequency times a number fixed by the symmetry energy, in keeping
with the empirical relation of Ref.~\cite{SotaniTogashi26}, and it falls by a
quarter to a third along an early-onset quarkyonic one. Of the three
modes, only the $g$ mode carries a scale-free signal of composition, and that
is the sharpest statement we can make of what a $g$ mode is for.

The paper is organised as follows. Section~\ref{s:form} sets out the model,
derives the two sound speeds and states the oscillation problem;
Sec.~\ref{s:num} describes the numerical implementation and the verification
record; Sec.~\ref{s:res} presents the results; Sec.~\ref{s:disc} contrasts them
with the Gibbs mixed-phase case and scopes the limitations; and
Sec.~\ref{s:sum} offers our conclusions.

\section{Formalism}
\label{s:form}

The calculation has three parts, and only the second is new. Section
\ref{s:eos} fixes the equation of state, following Zhao and
Lattimer~\cite{Zhao20}. Sections \ref{s:frozen} to \ref{s:quad} derive the
adiabatic sound speed of quarkyonic matter, which has not been done before and
which is where the physics of this paper lives: the outcome is
Eq.~(\ref{e:quad}), a closed expression for $\cs^2-\ce^2$ from which the quarks
have disappeared. Sections \ref{s:osc} and \ref{s:damp} then state the
oscillation problem and the radiated observables in the standard way, and
Sec.~\ref{s:assume} collects every approximation in one place.

\subsection{Quarkyonic matter in $\beta$ equilibrium}
\label{s:eos}

Matter consists of interacting nucleons ($n,p$), non-interacting deconfined
quarks ($d,u$), and leptons ($e^-,\mu^-$). Above a transition density $n_t$ the
nucleons occupy a momentum shell, so that with $g_s=2$ and $N_c=3$
\begin{equation}
  n_{n,p}=\frac{k_{F(n,p)}^3-k_{0(n,p)}^3}{3\pi^2},\quad
  n_{d,u}=\frac{k_{F(d,u)}^3}{\pi^2},\quad
  n_{e,\mu}=\frac{k_{F(e,\mu)}^3}{3\pi^2}.
  \label{e:dens}
\end{equation}
Following Eq.~(17) of Ref.~\cite{Zhao20} the lower edge of the nucleon shell is
a function of the corresponding Fermi momentum alone,
\begin{equation}
  k_{0i}=\left(k_{Fi}-k_{ti}\right)
  \left[1+\frac{\Lambda^2}{(\hbar c)^2 k_{Fi}k_{ti}}\right],
  \qquad i=n,p,
  \label{e:k0}
\end{equation}
with $k_{ti}$ the Fermi momentum of species $i$ at the onset and $\Lambda$ a
parameter controlling how quickly the shell opens. Three distinct quantities
in this paper are written with some form of the letter $\Lambda$, and we fix
the convention here: an unadorned $\Lambda$ is always this shell scale, in
MeV; $\Lambda_{\rm QCD}$ appears only in the strong-interaction timescale of
Sec.~\ref{s:frozen}; and the dimensionless tidal deformability is always
barred, $\Lbar$. Where a model is labelled by a superscript, as in
QY$_{50}^{800}$, the superscript is $\Lambda$ in MeV. The exponent on $\Lambda$ is
two, not the three of the original chargeless construction of McLerran and
Reddy~\cite{McLerran19}: Ref.~\cite{Zhao20} makes that replacement deliberately,
to slow the opening of the shell, and abandons with it the locking condition
$k_{0n}=N_ck_{Fd}$, which is incompatible with minimising the energy over the
composition. Fixing $(\Lambda,n_t)$ fixes the McLerran--Reddy shell parameter
$\kappa$ of Ref.~\cite{McLerran19} for each species,
\begin{equation}
  \kappa_i=N_c^2\left(\frac{\hbar c\,k_{ti}}{\Lambda}
  -\frac{\Lambda}{\hbar c\,k_{ti}}\right),
  \label{e:kappa}
\end{equation}
which we quote in Sec.~\ref{s:num} as one of the checks against the published
tables, and which will reappear in Sec.~\ref{s:noB}: $\kappa_i<0$ is precisely
the condition under which the nucleon shell has a finite capacity.
Baryon number and charge
neutrality read
\begin{equation}
  \nB=n_n+n_p+\tfrac13(n_d+n_u),\qquad
  n_p+\tfrac13(2n_u-n_d)=n_e+n_\mu .
  \label{e:cons}
\end{equation}
Two \emph{strong} conditions equilibrate quarks with nucleons,
\begin{equation}
  \mu_d=\tfrac13(2\mu_n-\mu_p),\qquad
  \mu_u=\tfrac13(2\mu_p-\mu_n),
  \label{e:strong}
\end{equation}
and two \emph{weak} conditions give $\beta$ equilibrium,
\begin{equation}
  \mu_e=\mu_\mu=\mu_n-\mu_p=\mu_d-\mu_u ,
  \label{e:weak}
\end{equation}
the last equality following from Eqs.~(\ref{e:strong}), and worth recording
because it is the statement that the quark sea inherits its weak imbalance
from the nucleons. The constituent quark masses are fixed, not free:
requiring both flavours to appear at $n_t$ gives
$m_d=\frac13(2\mu_{tn}-\mu_{tp})$ and $m_u=\frac13(2\mu_{tp}-\mu_{tn})$.

The nucleon interaction energy is the Zhao--Lattimer parameterisation
\begin{align}
  \frac{\varepsilon_{\rm int}}{n_s} &= 4x(1-x)\left(a_0u^2+b_0u^{\gamma+1}\right)
  \nonumber\\ &\quad
  +(1-2x)^2\left(a_1u^2+b_1u^{\gamma_1+1}\right),
  \label{e:zl}
\end{align}
with
\begin{equation}
  u=\frac{n_n+n_p}{n_s},\qquad x=\frac{n_p}{n_n+n_p},\qquad n_s=0.16\fm .
  \label{e:ux}
\end{equation}
Equation~(\ref{e:zl}) is $u$ times the potential energy \emph{per nucleon} of
Eq.~(10) of Ref.~\cite{Zhao20}, so every exponent here is one higher than in
theirs. Note that $u$ is built from the nucleon density and not from $\nB$:
above the transition the two diverge, because the nucleon shell saturates while
the quark sea keeps growing. For the baseline at $\nB=0.9\fm$ the nucleon
density has reached $1.95\,n_s$ while $\nB$ is $5.65\,n_s$. The nucleons
interact with each other, so it is the former that enters
Eq.~(\ref{e:zl}). The sub-parameters
$(a_0,b_0,\gamma,a_1,b_1,\gamma_1)$ are fixed by the saturation point
($B=16\mev$, $K=220\mev$ for symmetric matter, $S_v=31\mev$), the chosen slope
$L$ and $\gamma_1=5/3$. We take a grid of ten EOSs: seven quarkyonic sets varying
$L\in\{30,50,70\}\mev$, $\Lambda\in\{800,1100,1400\}\mev$ and
$n_t\in\{0.3,0.5\}\fm$, and the three purely hadronic ZL EOSs with the same $L$
values, which serve as the controls. The set $L=50\mev$, $\Lambda=1400\mev$,
$n_t=0.3\fm$ is the baseline. Below $\nB=0.08\fm$ the core table is spliced to
the SLy crust in the piecewise-polytropic representation of Ref.~\cite{Read09}.

\subsection{The two sound speeds}
\label{s:frozen}

\emph{Why should a star have two sound speeds at all?} The distinction is
standard in the $g$-mode literature~\cite{Reisenegger92,Lai94}, but everything
in this paper follows from the gap between them, so it is worth setting out
concretely.

Imagine taking a small parcel of neutron-star matter and squeezing it. The
pressure rises, and how much it rises depends on what the parcel is permitted
to do while you squeeze. If you squeeze slowly enough that every reaction inside
the parcel has time to run to completion, the parcel stays in full chemical
equilibrium and its composition is always the one the star would have chosen at
that density; the pressure change is then $dp=\ce^2\,d\varepsilon$ with
\begin{equation}
  \ce^2=\left(\frac{dp}{d\varepsilon}\right)_{\beta\text{-eq}},
\end{equation}
the slope of the $p(\varepsilon)$ curve tabulated for the star. If instead you
squeeze quickly, the slow reactions do not have time to respond, the parcel
keeps the composition it started with, and the pressure change is governed by a
different derivative, $\cs^2$. A parcel carrying a composition inherited from
somewhere else is out of chemical equilibrium with its new surroundings, and
being out of equilibrium costs energy. That extra cost makes the parcel
slightly stiffer than its background, so that $\cs>\ce$~\cite{Reisenegger92},
provided the
equilibrium composition really is the state of lowest energy at that density.
That proviso is not decoration: Eq.~(\ref{e:quad}) below turns it into an
explicit positive-definiteness condition. No equation of state on our grid
violates it; the closest approach is the $L=30\mev$ hadronic control, whose
leptons are squeezed out entirely above $\nB\simeq0.79\fm$, leaving pure
neutron matter that is exactly barotropic, so that $\cs=\ce$ and the buoyancy
vanishes rather than changing sign. Wherever the strict ordering holds, the
restoring force of a $g$ mode is built out of nothing but this difference: set
$\cs=\ce$ and the mode ceases to exist.

Deciding which reactions count as fast and which as slow is therefore not a
detail of the calculation; it \emph{is} the calculation. The $g$-mode periods
computed here run from $1.9$ to $12.4$~ms ($81$ to $522$~Hz above
$1.0\,\Msun$). Set that against the two families of reaction available to
quarkyonic matter.

The quark--nucleon conditions, Eqs.~(\ref{e:strong}), are maintained by the
strong interaction. They are the equilibrium conditions for
$n\leftrightarrow u\,d\,d$ and $p\leftrightarrow u\,u\,d$, that is for the
transfer of quarks between the confined population in the momentum shell and
the deconfined sea. Both processes conserve flavour and both proceed on the
strong-interaction timescale $\hbar/\Lambda_{\rm QCD}\sim10^{-23}$~s, some
twenty-one orders of magnitude shorter than the mode period. On the timescale
of an oscillation these conditions are not approximately satisfied; they are
exact. Note that no flavour is changed anywhere in this argument: turning a $d$
quark into a $u$ quark requires the weak interaction and belongs to the slow
family below.

The reactions that change the lepton content ($n\to p+e^-+\bar\nu_e$, $d\to u+e^-+\bar\nu_e$, and $\mu^-\to e^-+\bar\nu_e+\nu_\mu$) are weak. In a
cold catalysed star the modified-Urca rate gives relaxation times of seconds to
hours~\cite{Reisenegger92}, many orders of magnitude \emph{longer} than the
period; Refs.~\cite{Andersson19,Counsell24} treat the finite rate explicitly
and Sec.~\ref{s:fast} bounds what it would do here. On the timescale
of an oscillation the lepton fractions do not move at all.

So the parcel carries two labels that stay fixed while it is displaced, the
lepton fractions
\begin{equation}
  x_e=n_e/\nB,\qquad x_\mu=n_\mu/\nB,
\end{equation}
and everything else re-equilibrates instantly. The adiabatic sound speed is
therefore
\begin{equation}
  \cs^2=\left(\frac{\partial p}{\partial\varepsilon}\right)_{x_e,x_\mu},
  \label{e:cs2def}
\end{equation}
with Eqs.~(\ref{e:cons}), and (\ref{e:strong}) re-imposed at every step of the
derivative. The counting works out: at fixed $(x_e,x_\mu)$ the four unknowns
$(n_n,n_p,n_d,n_u)$ are fixed by two conservation laws and two
strong-equilibrium conditions, so the displaced state is unique and the
derivative is well defined.

Two familiar limits are recovered. Delete the quarks and
Eq.~(\ref{e:cs2def}) reduces to ``hold the proton fraction fixed'', the
standard $npe\mu$ prescription of Reisenegger and
Goldreich~\cite{Reisenegger92}. Delete the muons as well and one recovers the
$npe$ case of Lai~\cite{Lai94}. What is new is the case in between, in which
one set of chemical conditions is fast and another slow; it is not obtained by
freezing every fraction at once.

\subsection{Why the conventional prescription is unavailable}
\label{s:noB}

The hybrid-star literature takes the adiabatic derivative at fixed particle
fractions: every $x_i=n_i/\nB$ is pinned at its equilibrium value and all
densities scale together, on the grounds that no reaction is fast enough to
change them~\cite{Jaikumar21,ZhaoQNM22}. There is a physical objection to
importing that recipe here: the quark--nucleon conditions are
strong-interaction conditions, and no sense can be made of calling them slow.
But even setting the objection aside, the recipe collides with an obstruction
of a more stubborn kind. It becomes singular in exactly the region where the
quarkyonic mechanism does its work.

The obstruction is kinematic. A nucleon species confined to the shell
$k_{0i}\le k\le k_{Fi}$ has density $n_i=(k_{Fi}^3-k_{0i}^3)/3\pi^2$ with
$k_{0i}$ given by Eq.~(\ref{e:k0}). From that equation
\begin{equation}
  k_{Fi}-k_{0i}=k_{ti}-\frac{\Lambda^2}{(\hbar c)^2k_{ti}}
  \left(1-\frac{k_{ti}}{k_{Fi}}\right),
  \label{e:width}
\end{equation}
so the shell width starts at $k_{ti}$ and decreases monotonically towards
$k_{ti}-\Lambda^2/[(\hbar c)^2k_{ti}]$. If that limit is negative the width
reaches zero at a finite $k_{Fi}$: the shell closes, $n_i$ returns to zero, and
since $n_i$ was rising at the onset it must pass through a maximum in between.
The condition is
\begin{equation}
  \Lambda>\hbar c\,k_{ti}.
  \label{e:ceil}
\end{equation}
If instead $\Lambda<\hbar c\,k_{ti}$ the width tends to a positive constant,
$n_i$ grows without bound, and there is no ceiling. By Eq.~(\ref{e:kappa}), the
condition (\ref{e:ceil}) is simply $\kappa_i<0$, so whether the shell saturates
can be read straight off the McLerran--Reddy parameter: the baseline set has
$\kappa_n=-29.0$ and $\kappa_p=-74.2$, both negative.
For $n_t=0.3\fm$ this threshold is $\hbar c\,k_{tn}=399\mev$. Below it the
shell never closes and the frozen-composition prescription remains well
defined; above it the neutron shell has a finite capacity, which for the
baseline $\Lambda=1400\mev$ is $n_n^{\rm max}=0.292\fm$ and which falls with
increasing $\Lambda$ ($0.346\fm$ at $800\mev$, $0.304\fm$ at $1100\mev$).

This threshold costs us nothing in practice, because the interesting part of
the parameter space lies entirely above it. Quarkyonic matter is invoked to
support a two-solar-mass star, and the stiffening is controlled by $\Lambda$:
at $L=50\mev$ and $n_t=0.3\fm$ we obtained $M_{\rm max}=1.73\,\Msun$ for
$\Lambda=400\mev$, $2.00\,\Msun$ for $600\mev$ and $2.12\,\Msun$ for
$700\mev$, so the mass of PSR J0740+6620 requires $\Lambda\gtrsim650\mev$, comfortably
inside the region where Eq.~(\ref{e:ceil}) holds. Those three
points are quoted only to locate the threshold and lie below the range
$\Lambda=800$--$1700\mev$ surveyed in Ref.~\cite{Zhao20}, which notes that for
$\Lambda\lesssim500\mev$ quarkyonic matter requires an implausibly low
transition density; our production grid stays inside their range. Wherever the
construction does the job it was invented for, the shell has a ceiling.

And the ceiling, once it exists, is reached. At the centre of the
maximum-mass configuration the neutron shell holds $99.8$--$100.0\%$ of its
capacity in every one of the seven quarkyonic sets; at the centre of a
$1.4\,\Msun$ star it holds $96$--$100\%$ in six of them and $86\%$ in the
seventh, the $L=70\mev$, $n_t=0.5\fm$ set, whose core has only just crossed
$n_t$ at that mass. That is the content of the quarkyonic
mechanism: the shell is full and every further unit of baryon number goes into
the quark sea. Compressing a fluid element at fixed particle fractions then
demands of the shell a density it barely possesses, and the frozen-composition
compressibility diverges as the ceiling is approached. Prescription
(\ref{e:cs2def}) is thus not one option among several. It is the physically
correct one, because the quark--nucleon conditions are strong-interaction
conditions; and it is the only one that remains regular precisely where
quarkyonic matter earns its keep.

A question of degree remains. \emph{How much would our answer move if the
split between the two sound speeds were larger or smaller than we compute?}
Writing $\cs^2=\ce^2(1+\varsigma)$, the buoyancy factor is
$\ce^{-2}-\cs^{-2}=\ce^{-2}\varsigma/(1+\varsigma)$, which is linear in
$\varsigma$ to relative accuracy $\varsigma$ itself. On this grid $\varsigma$
stays below $0.11$
everywhere and below $0.062$ for the baseline set, so replacing $\cs^2$ by
$\ce^2+\delta(\cs^2-\ce^2)$ should give $f_{g_1}\propto\sqrt{\delta}$ to a few
per cent. We verified this directly rather than relying on the estimate: at
$1.4\,\Msun$ the baseline quarkyonic star gives
$98.5$, $138.5$ and $193.3$~Hz for $\delta=0.5$, $1$ and $2$, within $1.3\%$ of
exact square-root scaling, and the hadronic control behaves identically.
Reference~\cite{SotaniTogashi26} fits the same sensitivity for nucleonic
equations of state and obtains $-28\%$ and $+38\%$ at $\delta=0.5$ and $2$,
against our $-29\%$ and $+40\%$. A factor-of-two error in the sound-speed split
would move $f_{g_1}$ by about $40\%$, which is small compared with the
factor-of-two suppression and the sign reversal reported below.

\subsection{The quarks drop out, and what is left is a quadratic form}
\label{s:quad}

What does a change of composition cost in energy? This is the quantity out of
which the buoyancy is built, and the answer turns out to be far simpler than
the tally of six species would suggest.

Start from the general variation
$d\varepsilon=\sum_i\mu_i\,dn_i$ over all six species, impose baryon number
conservation and charge neutrality, Eqs.~(\ref{e:cons}), and use the strong
conditions (\ref{e:strong}) to trade $\mu_d$ and $\mu_u$ for $\mu_n$ and
$\mu_p$. Grouping the quark terms with the nucleon they are slaved to,
\begin{align}
  \sum_{i=n,p,d,u}\!\!\mu_i\,dn_i
  &=\mu_n\!\left[dn_n+\tfrac13(2dn_d-dn_u)\right]
  \nonumber\\ &\quad
   +\mu_p\!\left[dn_p+\tfrac13(2dn_u-dn_d)\right]
  \nonumber\\
  &=\mu_n\,d\nB-(\mu_n-\mu_p)(dn_e+dn_\mu),
\end{align}
where the last line uses $\nB=n_n+n_p+(n_d+n_u)/3$ and
$n_p+\tfrac13(2n_u-n_d)=n_e+n_\mu$. Restoring the lepton terms
$\mu_e\,dn_e+\mu_\mu\,dn_\mu$ gives
\begin{equation}
  d\varepsilon=\mu_n\,d\nB+\mut_e\,dn_e+\mut_\mu\,dn_\mu,
  \qquad
  \mut_i\equiv\mu_i-(\mu_n-\mu_p).
  \label{e:master}
\end{equation}

Look at what is missing. Six species went in and the quarks came out. The
energy cost of any change of composition is carried entirely by $\mu_n$ and by
the two \emph{chemical imbalances} $\mut_e$ and $\mut_\mu$, each measuring how
far the matter sits from weak equilibrium in one lepton channel. The reason is
physical rather than algebraic. A quark in this construction is not an
independent thermodynamic degree of freedom: its chemical potential is
determined by the nucleons through Eq.~(\ref{e:strong}), so when you add a
$d$ quark you are not making a free choice, you are making a bookkeeping entry
against $\mu_n$ and $\mu_p$. The quark sea is slaved to the nucleon shell.
Equation~(\ref{e:master}) is therefore the quarkyonic counterpart of the
standard $npe\mu$ relation, and it has exactly the same form, with the same two
imbalances: a considerable simplification, and the structural result on
which everything below rests.

Two immediate consequences. In full $\beta$ equilibrium the imbalances vanish
by definition, so $d\varepsilon/d\nB=\mu_n$, which integrates to the familiar
$\varepsilon+p=\mu_n\nB$ and gives
\begin{equation}
  \ce^2=\frac{1}{\mu_n}\frac{dp}{d\nB}=\frac{d\ln\mu_n}{d\ln\nB}.
  \label{e:ce2}
\end{equation}
Away from equilibrium but at fixed lepton fractions,
$(\partial\varepsilon/\partial\nB)_x=\mu_n+x_e\mut_e+x_\mu\mut_\mu
=(\varepsilon+p)/\nB$, so
$\cs^2=\nB(\varepsilon+p)^{-1}(\partial p/\partial\nB)_x$. Both speeds are
evaluated on the same $\beta$-equilibrated background star, where the two
prefactors are equal; the difference between them is a difference of
derivatives, not of states.

Now take the difference. Treat $\varepsilon$ as a function of the three
variables $(\nB,x_e,x_\mu)$. Equation~(\ref{e:master}) says
$(\partial\varepsilon/\partial x_i)_{\nB}=\nB\mut_i$: the imbalance is the
energy cost per unit lepton fraction, which is the statement that
$\beta$ equilibrium minimises the energy at fixed density. Using
$p=\nB(\partial\varepsilon/\partial\nB)_x-\varepsilon$ and the equality of
mixed partial derivatives,
\begin{equation}
  \left(\frac{\partial p}{\partial x_i}\right)_{\nB}
  =\nB^2\left(\frac{\partial\mut_i}{\partial\nB}\right)_{x}.
\end{equation}
Expand $dp/d\nB$ along the equilibrium sequence, where $x_e$ and $x_\mu$ both
change with density, and subtract Eq.~(\ref{e:ce2}):
\begin{equation}
  \cs^2-\ce^2=-\frac{\nB^2}{\mu_n}\sum_{i=e,\mu}
  \left(\frac{\partial\mut_i}{\partial\nB}\right)_x\frac{dx_i}{d\nB}.
  \label{e:diff}
\end{equation}
The two speeds differ only because the equilibrium composition changes with
depth. In a star of uniform composition they would be identical and there would
be no $g$ mode, which is the standard statement that a barotrope has no
buoyancy.

Equation~(\ref{e:diff}) still contains the composition gradients
$dx_i/d\nB$, which are properties of the background rather than of the parcel.
They can be eliminated. Differentiating the equilibrium conditions
$\mut_e=\mut_\mu=0$ along the sequence gives
$\bm A+\bm M\,(dx_e/d\nB,\,dx_\mu/d\nB)^{\!\top}=0$, where
\begin{equation}
  A_i=\left(\frac{\partial\mut_i}{\partial\nB}\right)_x,
  \qquad
  M_{ij}=\left(\frac{\partial\mut_i}{\partial x_j}\right)_{\nB},
\end{equation}
and substituting back,
\begin{equation}
  \boxed{\;\cs^2-\ce^2=\frac{\nB^2}{\mu_n}\,
  \bm{A}^{\!\top}\bm{M}^{-1}\bm{A}\;}
  \label{e:quad}
\end{equation}

Every object in Eq.~(\ref{e:quad}) has a plain physical meaning, and the
buoyancy emerges from a competition between two of them. The vector $\bm A$
measures how hard compression drives the matter out of weak equilibrium:
squeeze at fixed lepton fractions and the imbalances grow at the rate $A_i$.
The matrix $\bm M$ measures how stiffly the energy resists a change of lepton
fraction at fixed density; it is the Hessian of $\varepsilon$ in the
weak-composition directions,
$M_{ij}=\nB^{-1}\partial^2\varepsilon/\partial x_i\partial x_j$. A large
driving against a soft restoring energy gives a large buoyancy; a weak driving
against a stiff one gives almost none. Ultimately then, the entire physics of
the $g$ mode in quarkyonic matter reduces to the question of which of these two
factors the quarks contrive to change.

Two properties of Eq.~(\ref{e:quad}) are worth recording. First, whenever the
$\beta$-equilibrium point is a genuine minimum of the energy at fixed $\nB$,
$\bm M$ is positive definite, the quadratic form is non-negative and
$\cs^2\ge\ce^2$, so $N^2\ge0$ and the matter is stable against
convection~\cite{Finn87,Lai94}. The
formula therefore doubles as a stability diagnostic, and we use it as one; on
this grid it never fires, the marginal case being the barotropic deep core of
the $L=30\mev$ control discussed in Sec.~\ref{s:lim}.
Second, it reduces correctly. Drop the muons and it collapses to
$\cs^2-\ce^2=\nB^2A_e^2/(\mu_n M_{ee})$, the $npe$ result of
Ref.~\cite{Lai94}; drop the quarks as well and $\bm A$ and $\bm M$ reduce to
derivatives of the symmetry energy and one recovers the $npe\mu$ expressions
of Ref.~\cite{Reisenegger92}.

The quarks contribute no term of their own to Eq.~(\ref{e:quad}), there is no $x_d$ or $x_u$ anywhere in it, but they are not passive. They enter
through $\bm A$ and $\bm M$, and they enter strongly. When the matter is compressed at
fixed lepton fractions, the quark densities readjust through
Eq.~(\ref{e:strong}), and absorb part of the change in $\mu_n-\mu_p$ that would
otherwise show up as an imbalance. The quark sea thus buffers the very quantity that drives the buoyancy.
Section~\ref{s:res} quantifies how much.

\subsection{Stellar models and non-radial oscillations}
\label{s:osc}

Stellar models follow from the Tolman--Oppenheimer--Volkoff
equations~\cite{Tolman39,Oppenheimer39} for the metric
$ds^2=e^{\nu}dt^2-e^{\lambda}dr^2-r^2d\Omega^2$, integrated from a central
pressure to the surface, with $\nu$ normalised to the exterior Schwarzschild
value at $R$. The dimensionless tidal deformability
$\Lbar=\frac23k_2C^{-5}$, with $C=GM/Rc^2$, is obtained from the standard $l=2$
static perturbation equation for $y=rH'/H$~\cite{Hinderer08}.

For the oscillations we work in the relativistic Cowling
approximation~\cite{Cowling41}, which holds the metric fixed while the fluid
oscillates. It is the better approximation for $g$ modes than for the $f$
mode, and for the same reason in both cases: a low-frequency, nearly
divergence-free displacement couples weakly to the spacetime, while the $f$
mode is a large-scale density oscillation that does not. The two errors also
have opposite signs. For $g$ modes Cowling \emph{underestimates} the
frequency, by up to $10\%$ and by more at high
mass~\cite{ZhaoQNM22,SotaniTogashi26}; Refs.~\cite{Counsell25,Kruger25} do not
quantify the shift but argue from the same weak gravitational-wave coupling
that the approximation is adequate for $g$ modes. For the $f$ mode it
\emph{overestimates} the frequency, by up to $30\%$ in the surveys of
Refs.~\cite{Yoshida97,Chirenti15} and by about $15\%$ in the models of
Ref.~\cite{Counsell25}; our own stars give $5.8$--$16.2\%$
(Sec.~\ref{s:num}). We did not solve the full general-relativistic problem, so
these are the error estimates the paper carries: every $g$-mode frequency
below is a lower bound at the ten-per-cent level, and we return in
Sec.~\ref{s:sum} to what a correction of that size does to the conclusions.

We integrate the first-order system of Ref.~\cite{Jaikumar21}, in the variables
\begin{equation}
  U=r^2e^{\lambda/2}\xi^r,\qquad
  \mathcal{V}=\frac{\delta p}{\varepsilon+p},
  \label{e:UV}
\end{equation}
with $\xi^r$ the radial displacement and $\delta p$ the Eulerian pressure
perturbation, both carrying the factor $Y^l_me^{i\omega t}$. Writing
$g\equiv-p'/(\varepsilon+p)=\nu'/2$ for the local gravity, the equations are
\begin{align}
  \frac{dU}{dr} &= \frac{g}{\cs^2}\,U
  +e^{\lambda/2}\left[\frac{l(l+1)e^{\nu}}{\omega^2}-\frac{r^2}{\cs^2}\right]\mathcal{V},
  \label{e:U}\\
  \frac{d\mathcal{V}}{dr} &= e^{\lambda/2-\nu}\left(\omega^2-N^2\right)\frac{U}{r^2}
  +g\left(\frac{1}{\ce^2}-\frac{1}{\cs^2}\right)\mathcal{V},
  \label{e:V}
\end{align}
with the relativistic \bv\ frequency
\begin{equation}
  N^2=g^2\left(\frac{1}{\ce^2}-\frac{1}{\cs^2}\right)e^{\nu-\lambda}.
  \label{e:N2}
\end{equation}
The regular solution at the centre is $U=r^{l+1}$ and
$\mathcal{V}=e^{-\nu(0)}\omega^2r^l/l$, fixing the arbitrary overall amplitude, and the surface condition is a
vanishing Lagrangian pressure perturbation,
\begin{equation}
  \Delta p=(\varepsilon+p)\left(\mathcal{V}-g\,\xi^r\right)=0
  \qquad\text{at }r=R .
  \label{e:dplag}
\end{equation}
Equations~(\ref{e:U})--(\ref{e:V}) are linear in $(U,\mathcal{V})$ and their
coefficients depend on $\omega^2$ only through explicit factors, which is what
makes the vectorised frequency scan of Sec.~\ref{s:num} possible.

The energy and radiation integrals below are conventionally written in Thorne's
perturbation functions~\cite{Thorne67},
\begin{equation}
  \xi^r=r\,e^{-\lambda/2}W\,Y^l_m e^{i\omega t},\qquad
  \xi^\theta=-V_T\,\partial_\theta Y^l_m e^{i\omega t},
\end{equation}
which are recovered algebraically from the integrated solution,
\begin{equation}
  W=\frac{U}{r^3},\qquad
  V_T=-\frac{e^{\nu}\mathcal{V}}{\omega^2r^2}.
  \label{e:toW}
\end{equation}
Following Refs.~\cite{Jaikumar21,Constantinou21,ZhaoQNM22,Zheng23} we set
$\cs=\ce$ and $N=0$ in the crust, so the modes reported here are core modes.
We solve for $l=2$; the eigenvalues order as
$\omega_{g_n}<\dots<\omega_{g_1}<\omega_f<\omega_{p_1}<\dots$, with $n$ the
number of nodes of $W$.

\subsection{Damping, strain and the tidal resonance}
\label{s:damp}

The GW damping time is $\taugw=2E/P_{\rm GW}$, where the mode energy
follows~\cite{McDermott83}
\begin{equation}
  \frac{dE}{dr}=\frac{\omega^2}{2}(p+\varepsilon)e^{(\lambda-\nu)/2}r^4
  \left[W^2+l(l+1)V_T^2\right],
  \label{e:dEdr}
\end{equation}
and the radiated power is estimated as~\cite{Reisenegger92}
\begin{equation}
  P_{\rm GW}=\frac{(l+1)(l+2)}{8\pi(l-1)l}
  \left[\frac{4\pi\omega^{l+1}}{(2l+1)!!}\int_0^R\!dr\,r^{l+2}\delta\varepsilon
  \right]^{2},
  \label{e:pgw}
\end{equation}
with $\delta\varepsilon$ the Eulerian energy-density perturbation. That
quantity deserves care, because the form in which it is usually written is
numerically treacherous, and because the way it is written here is where the
two sound speeds re-enter.

By the definition of $\cs$, a displaced element satisfies
$\Delta\varepsilon=\Delta p/\cs^2$, so
$\delta\varepsilon=\Delta p/\cs^2-\xi^r\,d\varepsilon/dr$. Writing
$\Delta p=\delta p+\xi^r dp/dr$, using $d\varepsilon/dr=(dp/dr)/\ce^2$ on the
$\beta$-equilibrated background and $dp/dr=-g(\varepsilon+p)$,
\begin{equation}
  \delta\varepsilon=(\varepsilon+p)\left[\frac{\mathcal{V}}{\cs^2}
  +g\,\xi^r\left(\frac{1}{\ce^2}-\frac{1}{\cs^2}\right)\right].
  \label{e:depsalg}
\end{equation}
The first term is the adiabatic compression of the element and the second the
advection of the background profile; the second vanishes wherever $\cs=\ce$,
so the same difference of sound speeds that supports the mode also controls
what it radiates. Equation~(\ref{e:depsalg}) is what we evaluate. It is
algebraic in the integrated $(U,\mathcal{V})$ and requires no numerical
differentiation.

The equivalent expression in Thorne's variables,
\begin{align}
  \delta\varepsilon = &-(p+\varepsilon)\left[e^{-\lambda/2}
  \left(3W+r\frac{dW}{dr}\right)+l(l+1)V_T\right]
  \nonumber\\ &
  -r\frac{d\varepsilon}{dr}e^{-\lambda/2}W ,
  \label{e:deps}
\end{align}
is the one usually quoted, but as noted already in Ref.~\cite{Reisenegger92}
its two terms have opposite signs over most of the star and cancel strongly,
which makes $P_{\rm GW}$ and hence $\taugw$ delicate to evaluate directly. We
use Eq.~(\ref{e:deps}) only as a check on Eq.~(\ref{e:depsalg}), which it
reproduces to $2\times10^{-14}$ (Sec.~\ref{s:num}).

The strain amplitude in the quadrupole approximation, for inclination
$\sin\alpha=1$, is $|h_+|=3|\ddot{Q}_{33}|/(2D)$ with
$|\ddot{Q}_{33}|=\frac43\sqrt{\pi/5}\,\omega^2\int dr\,r^4\delta\varepsilon$.
Since $E\propto A^2$ and $\delta\varepsilon\propto A$ for an overall amplitude
$A$, $\taugw$ is amplitude independent while $|h_+|\propto\sqrt{E}$; we
normalise to $E=10^{51}$~erg~\cite{Lugones21}. The minimum energy that must be
radiated for a detection at signal-to-noise ratio $S/N$
is~\cite{Kokkotas01,Andersson11}
\begin{equation}
  \frac{E_{\rm GW}}{\Msun}=3.5\times10^{36}\,\frac{1+4Q^2}{4Q^2}\,
  \frac{S_n}{1\,\mathrm{s}}
  \left(\frac{S}{N}\frac{D}{10\,\mathrm{kpc}}\frac{f}{1\,\mathrm{kHz}}\right)^2,
  \label{e:egw}
\end{equation}
with $Q=\pi f\taugw$ the quality factor.

In a binary inspiral the $l=m=2$ tidal field drives the star at twice the
orbital frequency, so a mode is resonantly excited when
$2\Omega_{\rm orb}=\omega_\alpha$, that is when the gravitational-wave
frequency equals the mode frequency,
\begin{equation}
  f_{\rm GW}^{\rm res}=f_\alpha .
  \label{e:res}
\end{equation}
At leading quadrupole order the time from that frequency to coalescence for a
binary of chirp mass $\mathcal{M}$ is
\begin{equation}
  t_{\rm coal}=\frac{5}{256}
  \left(\frac{G\mathcal{M}}{c^3}\right)^{-5/3}(\pi f_{\rm GW})^{-8/3}.
  \label{e:tcoal}
\end{equation}
Equations~(\ref{e:res}), and (\ref{e:tcoal}) involve only the mode frequency and
Kepler's law, and are therefore as robust as $f_{g_1}$ itself. The size of the
resulting phase shift additionally requires the tidal overlap integral, which
we do not quote here; see Sec.~\ref{s:noQ}.

\subsection{Assumptions}
\label{s:assume}

Every restriction on the calculation is collected here rather than left to be
discovered. Their consequences are quantified in Sec.~\ref{s:lim}.

\begin{enumerate}
\item \emph{Relativistic Cowling.} The metric is held fixed. We did not solve
  the full general-relativistic problem, so the size of the error is taken
  from the literature rather than from our own calibration: published
  comparisons put the $g$-mode frequency about ten per cent \emph{below} the
  full-GR value, the deficit growing with
  mass~\cite{ZhaoQNM22,SotaniTogashi26}. Our $f$-mode frequencies, which we
  can calibrate, lie $5.8$--$16.2\%$ above the full-GR fit of
  Ref.~\cite{Andersson98} (Sec.~\ref{s:num}). Every $g$-mode frequency quoted
  here is therefore a lower bound at the ten-per-cent level.
\item \emph{Zero temperature.} The equation of state, the composition and the
  buoyancy are $T=0$. Thermal buoyancy is absent, which is appropriate for a
  star older than a few minutes and not for a merger remnant.
\item \emph{No rotation.} The stars are spherical and non-rotating; rotational
  splitting and the Coriolis force are absent.
\item \emph{Frozen weak composition, instantaneous strong equilibrium.} On an
  oscillation period the lepton fractions do not change while quark--nucleon
  chemical equilibrium is re-imposed exactly. Section~\ref{s:fast} bounds what
  a finite weak rate would do.
\item \emph{Barotropic crust.} The spliced SLy crust carries no composition
  gradient, so $N^2=0$ there by construction. This is why no tidal overlap
  integral is quoted for the $g$ modes (Sec.~\ref{s:noQ}).
\item \emph{One construction, ten models.} All quarkyonic models use the
  Zhao--Lattimer momentum shell, scanned over $L$, $\Lambda$ and $n_t$. The
  grid spans a parameter range, not a range of constructions.
\item \emph{Bare nucleon masses.} The Zhao--Lattimer functional supplies no
  Landau effective mass, so none is used.
\end{enumerate}

\section{Numerical implementation and verification}
\label{s:num}

\subsection{Implementation}

The accuracy of the small quantity $\cs^2-\ce^2$ rests on three choices.

\emph{Two unknowns.} Every chemical potential is slaved to $(\mu_n,\mu_p)$: the
quark Fermi momenta follow from Eq.~(\ref{e:strong}), and the lepton momenta from
Eq.~(\ref{e:weak}), so the root find is over $(k_{Fn},k_{Fp})$ only, with the
two relations (\ref{e:cons}) as residuals. Threshold discontinuities (muon onset, quark onset) are thereby removed from the residual vector.

\emph{One stencil for both speeds.} $\ce^2$ and $\cs^2$ are evaluated by the
same central difference in $\nB$ at the same relative step, so their difference
is free of differencing bias and its sign is meaningful.

\emph{Vectorised mode scan.} Equations (\ref{e:U})--(\ref{e:V}) are linear and
their coefficients depend on $\omega^2$ only through explicit factors, so the
whole frequency scan is one vectorised fourth-order Runge--Kutta sweep, and
brackets are refined by vectorised subdivision. Because every $g$ mode of these
stars lies below $0.6$~kHz while the $f$ mode lies above $1.7$~kHz, the $f$ mode
is identified as the lowest root above $1$~kHz and its node count verified;
eigenfunctions are then integrated only for the six labelled modes.

\subsection{Verification}

The EOS layer reproduces the published Zhao--Lattimer sub-parameters exactly
($\gamma=1.2564$, $a_0=-129.27\mev$, $b_0=91.48\mev$, $a_1=-L/2-14.70\mev$,
$b_1=L/2-4.63\mev$), and, at the baseline transition, gives
$m_d=391.5\mev$, $m_u=240.5\mev$, $\kappa_n=-29.0$, $\kappa_p=-74.2$ against
the published $391.28$, $241.07$, $-29.00$, $-74.54$. The identity
$\varepsilon+p=\mu_n\nB$ holds to $<10^{-11}$ on every table. Most
importantly, the quadratic form (\ref{e:quad}), and a direct finite-difference
evaluation of $\cs^2-\ce^2$, two nearly disjoint code paths, agree to
$10^{-7}$--$10^{-4}$ relative.

For the oscillation layer we ran eleven gates, each with a tolerance fixed
before the run. The central limits $W\to1$ and $V_T\to-1/2$ are recovered to
$4\times10^{-7}$; the surface condition $\Delta p(R)=0$ holds to
$5\times10^{-11}$ of $\max|\Delta p|$; node counts satisfy $n_f=0$,
$n_{g_1}=n_{p_1}=1$, $n_{p_2}=2$; the mode-energy normalisation, the
invariance of $\taugw$ under a change of amplitude and the scaling
$|h_+|\propto\sqrt{E}$ all hold to machine precision; and doubling the radial
resolution shifts $f_{g_1}$ by $<10^{-6}$, $\taugw$ by $<0.3\%$ and $|h_+|$ by
$<0.2\%$. One tolerance was revised: the window on the Cowling $f$-mode excess
over the full-GR fit was declared as $15$--$40\%$ and widened to $0$--$45\%$
after the first run, because the excess at high mass came out smaller than we
had anticipated. The measured values are given below.

The check on $\delta\varepsilon$ needs to be described for what it is. Feeding
the right-hand sides of Eqs.~(\ref{e:U})--(\ref{e:V}) into Eq.~(\ref{e:deps})
collapses it algebraically onto Eq.~(\ref{e:depsalg}), so the agreement we
measure between them, $2\times10^{-14}$, tests that Eq.~(\ref{e:deps}) has
been transcribed correctly and nothing more. Evaluating Eq.~(\ref{e:deps})
independently, with $dW/dr$ and $d\varepsilon/dr$ taken by finite difference
rather than from the ODE, agrees with Eq.~(\ref{e:depsalg}) only to $42\%$ for
the baseline quarkyonic star and $57\%$ for its control, in the smooth core
away from the crust splice. That is the cancellation described below
Eq.~(\ref{e:deps}) doing its work, and it is the reason we evaluate
$\delta\varepsilon$ algebraically; but it means the radiated power carries no
tight internal check, and the external one, gate G11 below, is what bounds it:
the comparison of our $f$-mode damping times against the empirical fit of
Ref.~\cite{Andersson98}.

The strongest check is an independent second code path. Production uses the
$(U,\mathcal{V})$ system, Eqs.~(\ref{e:U})--(\ref{e:V}). Against it we
integrate the second-order Thorne system in the variables $(W,V_T)$ directly (different unknowns, different right-hand sides, a different surface condition, and only the background star in common), and recover the
eigenfrequency to $1.4\times10^{-6}$, $\taugw$ to $0.3\%$ and $|h_+|$ to
$0.15\%$. That is a test of the formalism, since the two systems are
algebraically equivalent but numerically unrelated. A separate
re-implementation of the $(U,\mathcal{V})$ system itself, with an independent
integrator and independent boundary handling, returns $f_f=2184.922$~Hz
against $2184.924$~Hz and $f_{g_1}=138.4364$~Hz against $138.4361$~Hz; that
one tests the integration rather than the formalism, and is quoted as such.

Finally, for four test configurations (the baseline quarkyonic model and its hadronic control, each at $1.4$ and $2.0\,\Msun$), the Cowling $f$-mode
frequency lies $13.2$, $5.8$, $16.2$ and $7.4\%$ above the
full-general-relativistic Andersson--Kokkotas fit~\cite{Andersson98}. The
sign is the expected one and the excess falls with mass. The same comparison tests the
normalisation of Eq.~(\ref{e:pgw}), which is otherwise checked only internally:
our $f$-mode damping times, $0.063$--$0.077$~s, are a factor $2.1$--$2.9$
shorter than Eq.~(8) of Ref.~\cite{Andersson98} gives
($0.147$--$0.227$~s). Almost all of that is the frequency
error propagating as $\omega^{6}$; dividing it out leaves a residual of
$0.60$--$0.92$, so the quadrupole estimate of the radiated power is low by
between $8$ and $40\%$ across the four configurations. Damping times in this paper should be read with that systematic
in mind. It cancels in the model-to-model comparisons, which are made at fixed
approximation. For the $g$ mode the error is smaller but not
negligible, and it has a definite sign. Zhao \emph{et al.}~\cite{ZhaoQNM22}
solved the same problem with the metric perturbations retained and found that
Cowling \emph{underestimates} the $g$-mode frequency by up to $10\%$, the
discrepancy growing with mass; Ref.~\cite{SotaniTogashi26} quotes below about
$10\%$ for cold stars, improving for lighter ones, and
Refs.~\cite{Counsell25,Kruger25} reach the same conclusion from the weakness of
the gravitational-wave coupling of a low-frequency mode. Every $g$-mode
frequency in this paper should therefore be read as a lower bound good to about
ten per cent, and every resonance frequency as one that shifts slightly upward
in full general relativity. None of our conclusions turns on that shift,
because all of them compare quarkyonic models with hadronic controls computed
in the same approximation. The Cowling error quoted for protoneutron
stars~\cite{SotaniTakiwaki20} is larger, but those stars are hot and
convective and the comparison does not carry over.

A direct comparison against published $g$-mode frequencies is possible for the
nucleonic case, and it is only partly satisfactory. Rebuilding the purely
hadronic Zhao--Lattimer equation of state at $L=55\mev$, the value used in
Table~II of Ref.~\cite{ZhaoQNM22}, we obtain $f_{g_1}=185.6$, $203.8$ and
$353.3$~Hz at $1.0$, $1.4\,\Msun$ and the maximum mass, which we place at
$2.02\,\Msun$. The Cowling column of Table~II of Ref.~\cite{ZhaoQNM22} gives $103$, $230$
and $311$~Hz at the same three points, with a maximum mass of
$2.11\,\Msun$.
The two calculations agree that $f_{g_1}$ rises monotonically with mass on a
nucleonic sequence, which is the property our central claim is contrasted
against, and they agree to $13\%$ at $1.4\,\Msun$ and at the maximum mass. They
disagree by close to a factor of two at $1.0\,\Msun$. The maximum masses also
differ, so the two equations of state are not identical despite the shared
label, and we have not been able to localise the remaining difference. We flag
it rather than leave it for a referee: the absolute nucleonic normalisation at
low mass is not settled, while the sign of $df_{g_1}/dM$, on which this paper
rests, is agreed.

One modelling choice deserves its own test. The SLy crust is joined to the
core at $n_{\rm cc}=0.08\fm$, and the $g$ mode has amplitude out to the
surface, so a referee is entitled to ask how much of $f_{g_1}$ is the splice.
Table~\ref{t:crust} rebuilds the baseline quarkyonic equation of state and its
hadronic control at $n_{\rm cc}=0.08$, $0.10$ and $0.12\fm$ and recomputes the
spectrum at fixed mass. The rebuild is independent of the production run and
reproduces its frequencies to $1\%$, which is itself a cross-check of the
solver. Across that range $f_{g_1}$ moves by $4.5\%$ at $1.4\,\Msun$
and $4.4\%$ at $2.0\,\Msun$ for the quarkyonic model, and by $2.9\%$ and
$0.1\%$ for the control; the $f$ mode moves by less than $0.5\%$ everywhere.
The crust convention is therefore worth a few per cent on the $g$-mode
frequency, an order of magnitude less than the factor-of-two suppression and
the change of sign in $df_{g_1}/dM$ that the quarkyonic transition produces. We
note that a realistic inner crust supports $g$ modes of its own, which can
undergo avoided crossings with the core modes and correlate with $L$ in their
own right~\cite{Sun25}; our crust is barotropic by construction, so those modes
are absent here and cannot contaminate the classification.

\begin{table}[t]
\caption{Sensitivity of the spectrum to the density $n_{\rm cc}$ at which the SLy crust is joined to the core. The last two columns give the largest fractional change in $f_{g_1}$ and in $f_f$ over the three splice densities, relative to the $0.08\fm$ value used throughout.}
\label{t:crust}
\begin{ruledtabular}
\begin{tabular}{lccccccc}
 & $M$ & \multicolumn{3}{c}{$f_{g_1}$ (Hz) at $n_{\rm cc}$ (fm$^{-3}$)} & \multicolumn{2}{c}{max.\ spread (\%)} \\
\cline{3-5}\cline{6-7}
Model & ($M_\odot$) & $0.08$ & $0.10$ & $0.12$ & $g_1$ & $f$ \\
\hline
QY$_{50}$ & 1.4 & 138.5 & 135.6 & 132.2 & 4.5 & 0.5 \\
QY$_{50}$ & 2.0 & 121.0 & 118.6 & 115.7 & 4.4 & 0.2 \\
\hline
ZL$_{50}$ & 1.4 & 160.1 & 158.0 & 155.4 & 2.9 & 0.5 \\
ZL$_{50}$ & 2.0 & 248.3 & 248.1 & 248.0 & 0.1 & 0.1 \\
\end{tabular}
\end{ruledtabular}
\end{table}

The tidal machinery discussed in Sec.~\ref{s:noQ} was subjected to four
additional analytic gates: for a Newtonian incompressible star the overlap
integral reproduces the closed form $3(2\pi)^{-1}\sqrt{2\pi/3}$ and the
mode-sum Love-number rule
$\sum_\alpha Q_\alpha^2/\bar\omega_\alpha^2=(2l+1)k_l/2\pi$ to one part in
$10^{10}$, and Radau's equation returns $k_2=3/4$ and $0.2599$ for the
incompressible and $n=1$ polytropic stars.

\section{Results}
\label{s:res}

\begin{figure*}[t]
\centerline{\includegraphics[width=\textwidth]{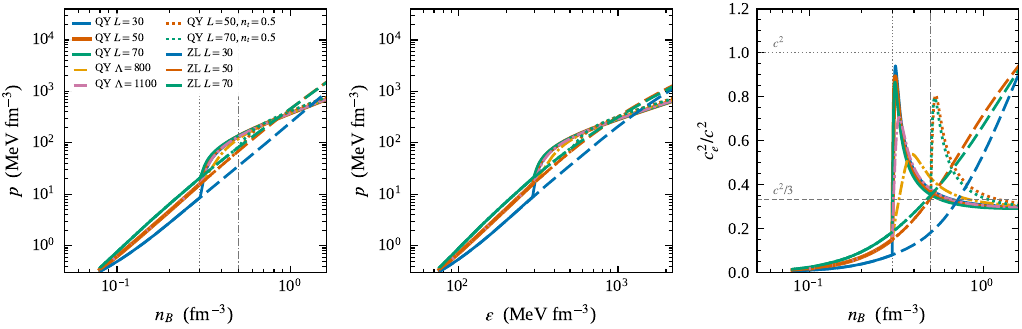}}
\caption{The ten equations of state used here. Left: pressure against baryon
number density, with the two transition densities marked. Centre: pressure
against energy density. Right: the squared equilibrium sound speed, with the
conformal value $c^2/3$ marked. Solid curves are quarkyonic with
$n_t=0.3\fm$, dotted with $n_t=0.5\fm$, dashed the purely hadronic
Zhao--Lattimer controls; colour encodes the symmetry-energy slope $L$ except
for the two curves that vary $\Lambda$ at fixed $L=50\mev$. Below
$n_B=0.08\fm$ every model carries the same SLy crust~\cite{Read09}.}
\label{f:eos}
\end{figure*}

\begin{figure*}[t]
\centerline{\includegraphics[width=\textwidth]{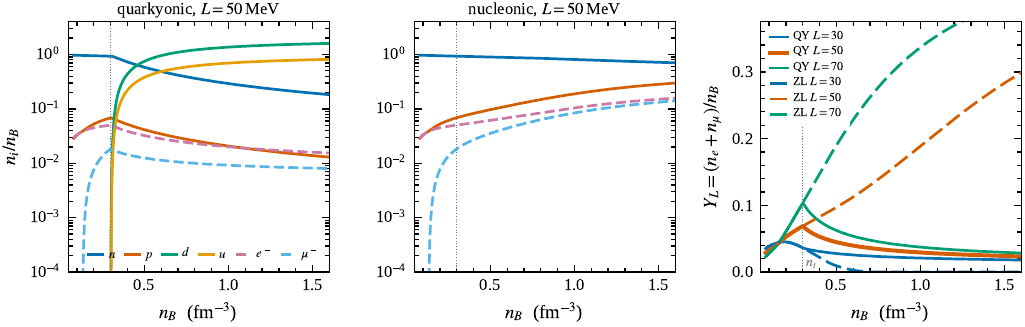}}
\caption{Composition of the matter. Left: particle fractions $n_i/\nB$ of the
baseline quarkyonic model; the quark fractions switch on at $n_t$ and the
$d$ quarks quickly become the most abundant species, while the neutron
fraction falls. Note that $n_d/\nB$ exceeds unity because $\nB$ counts baryon
\emph{number}, to which each quark contributes $1/3$. Centre: the purely
hadronic control with the same $L$, on the same scale. Right: the total lepton
fraction $Y_L=(n_e+n_\mu)/\nB$, which is the quantity whose gradient supplies
the entire buoyancy through Eq.~(\ref{e:quad}); curves as in
Fig.~\ref{f:eos}. Above $n_t$ the quarkyonic $Y_L$ turns over and
\emph{decreases}, while the hadronic $Y_L$ keeps rising. The hadronic
$L=30\mev$ curve falls to zero at $0.78\fm$ because its symmetry energy turns
over and the matter becomes pure neutron matter; above that density the model
is exactly barotropic, so $\cs=\ce$ and its buoyancy vanishes identically.}
\label{f:comp}
\end{figure*}

\begin{figure*}[t]
\centerline{\includegraphics[width=\textwidth]{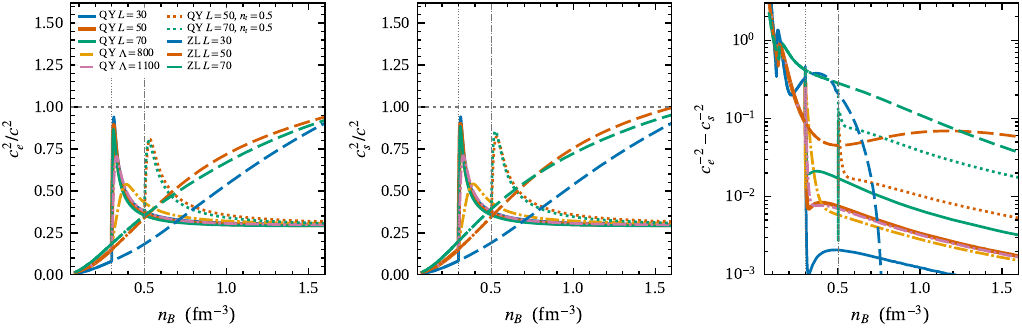}}
\caption{The two sound speeds and the buoyancy they produce; curves as in
Fig.~\ref{f:eos}. Left: the equilibrium speed $\ce^2$, the derivative of
$p(\varepsilon)$ along the $\beta$-equilibrium sequence. Centre: the adiabatic
speed $\cs^2$ of a displaced fluid element, Eq.~(\ref{e:cs2def}). Right: their
difference in the combination $\ce^{-2}-\cs^{-2}$ that sets $N^2$ through
Eq.~(\ref{e:N2}). The two speeds are so nearly equal that the left and centre
panels are hard to tell apart, which is the point: the buoyancy lives
entirely in the difference, and the buoyancy factor built from it drops by
about an order of magnitude across the transition in the early-onset models,
and by a factor of two to three in the late-onset ones. The step near $\nB\simeq0.134\fm$ in
the right panel is the muon threshold: the buoyancy factor roughly doubles,
from $0.47$ to $0.90$, when the second lepton channel opens, because a new compositional degree of
freedom is a new way for a displaced element to be out of equilibrium.}
\label{f:cs}
\end{figure*}

\subsection{The equation of state and its composition}

Figure~\ref{f:eos} shows the input. It is worth pausing over what the
quarkyonic construction actually does to it, because the mechanism is very
easily misread as a phase transition, which it is not.

Below $n_t$ the matter is ordinary nucleonic matter. At $n_t$ the quark sea
switches on, and from there upward the nucleons are excluded from the interior
of their own Fermi sphere: they are pushed into a shell
$k_{0i}\le k\le k_{Fi}$ whose lower edge rises with density. Because the
nucleons are forced to sit at high momentum, their kinetic energy per particle
rises faster than it would in a filled sphere, and the pressure rises with it.
That is the stiffening, and the left panel shows it as a kink in $p(\nB)$ over
an interval of roughly $0.02\fm$ at the fiducial $\Lambda=1400\mev$, widening
to about $0.12\fm$ at $\Lambda=800\mev$, steep but continuous, with no
plateau and no jump. On the $p(\varepsilon)$ plane in the centre panel the same feature is
barely visible at all, which is exactly why the mass--radius relation is a weak
discriminator: two equations of state can differ in how they got to a given
$p(\varepsilon)$ and produce nearly the same star.

The right panel is where the action is. The equilibrium sound speed overshoots
to
$0.94$, $0.89$ and $0.87$ at $L=30$, $50$ and $70\mev$ just above $n_t$, then
relaxes back towards the conformal value $c^2/3$ from above. A brief interval
of near-maximal stiffness followed by a return to $\sim c^2/3$ is precisely the
shape needed to reconcile a soft symmetry energy at saturation with a
two-solar-mass star, and it is the reason quarkyonic matter is of interest in
the first place. The height of that peak is what $\Lambda$ controls, and we scan it over the
range the Zhao--Lattimer construction is used in: $0.54$ at $\Lambda=800\mev$,
$0.71$ at $1100\mev$ and $0.89$ at the fiducial $1400\mev$. Every equation of
state on this grid is subluminal at every density realised inside a stable
star, the largest value being $\cs^2=0.97$ at the centre of the maximum-mass
ZL $L=30\mev$ configuration. (Two of the hadronic tables do cross $\cs^2=1$, at $\nB=1.63$ and $1.78\fm$,
but the maximum-mass stars built on them reach only $1.21$ and $1.09\fm$, so
the crossing lies outside every stable configuration.)

One feature of the right panel deserves emphasis because it is easy to miss
and it matters for everything that follows: the peak in $\ce^2$ is a
\emph{local} feature of the transition region, not a property of the dense
interior. By the centre of the maximum-mass star the quarkyonic models have
relaxed to $\ce^2=0.31$--$0.36$, that is to within $8\%$ of the conformal
value $c^2/3$, while the hadronic controls are still at
$0.72$--$0.97$ there. The stiffening that buys the two-solar-mass star is
spent in a narrow shell just above $n_t$; the deep core is soft and nearly
conformal. That is also the geometry the $g$ mode will report on, and it is
the reason the mode ends up confined to a shell rather than filling the
star.

The entire effect is compositional, so we should look at the composition
itself. Figure~\ref{f:comp} does exactly that.

The left panel shows what the quarkyonic transition does to the particle
content. Below $n_t$ the matter is $npe\mu$ with a proton fraction of a few per
cent, the familiar picture. At $n_t$ the $d$ and $u$ quarks appear and grow
rapidly; the $d$ quarks overtake the neutrons at $\nB=0.46\fm$ and the neutron
fraction then declines steadily. Charge neutrality then no longer
needs the electrons: the $d$ quarks carry the negative charge instead, and the
lepton fractions fall. Compare this with the hadronic control in the centre
panel, where the proton and lepton fractions climb monotonically because
nothing else is available to neutralise the matter.

The right panel is the one that matters for the $g$ mode. What enters
Eq.~(\ref{e:quad}) is not the composition itself but the way it changes with
depth, and the total lepton fraction $Y_L$ makes the contrast sharp. In a
purely hadronic star $Y_L$ rises steadily with density; the composition is
strongly stratified, the gradients $dx_i/d\nB$ are large, and the buoyancy is
correspondingly large. In a quarkyonic star $Y_L$ peaks at $n_t$, at $0.069$ for the baseline, and then falls, reaching $0.029$ by $\nB=1\fm$, where the hadronic control has
climbed to $0.189$.
Matter whose composition barely changes with depth is nearly unstratified, and
unstratified matter has no buoyancy at all. This is the same
statement as the collapse of $\ce^{-2}-\cs^{-2}$ in the next subsection, seen
one level closer to the microphysics.

\subsection{Where the buoyancy goes}
\label{s:chan}

Figure~\ref{f:cs} holds the mechanism, and it is not the mechanism that the
hybrid-star literature would lead one to expect.

The natural guess is that the quarks kill the $g$ mode by driving $\cs$ towards
$\ce$: with one chemical equilibrium now fast, surely the displaced parcel
simply re-equilibrates and stops pushing back. The guess is wrong. The left and
centre panels of Fig.~\ref{f:cs} show why: $\cs^2$ tracks $\ce^2$ closely above
the transition, but it tracked it just as closely below. We measure every change across the transition between two well-defined
densities: the last point below $n_t$, and the density at which $\ce^2$ reaches
its post-transition peak, which marks the end of the stiffening and moves with
$\Lambda$ (from $0.31\fm$ at $\Lambda=1400\mev$ to $0.39\fm$ at $800\mev$).
Between those two points the baseline separation goes from
$\cs^2-\ce^2=2.0\times10^{-3}$ to $6.5\times10^{-3}$, so the difference of the
squared speeds roughly triples. Across the grid it moves by a factor between
$0.47$ and $3.2$, in either direction. That is a modest change, and, as the next paragraph shows, an order of magnitude too small to account for what happens to the
buoyancy.

What changes, and changes with startling abruptness, is the \emph{product}. Across the
transition $\ce^2\cs^2$ jumps from $0.023$ to $0.80$, because both speeds jump
together: the shell structure of Eq.~(\ref{e:k0}) makes the matter very stiff
very quickly. That matters because the buoyancy does not depend on the
difference alone. Writing
\begin{equation}
  \frac{1}{\ce^2}-\frac{1}{\cs^2}=\frac{\cs^2-\ce^2}{\ce^2\cs^2},
  \label{e:ratio}
\end{equation}
the combination that enters $N^2$ is a difference divided by a product, and it
is the denominator that moves. For the baseline the numerator triples while
the denominator rises by a factor of $35$, so the buoyancy factor falls from
$0.088$ to $0.0080$, a factor of $11$, over a density interval of
$0.012\fm$. The right panel of Fig.~\ref{f:cs} shows the drop directly.
The same competition holds across the five early-onset sets, always with the
denominator winning: there the product rises by factors of $13$ to $133$ while
the difference moves by a factor between $0.47$ and $3.2$. In the two
$n_t=0.5\fm$ sets the product rises by only $4.9$ and $5.7$. How far the
buoyancy then falls depends on where the transition sits. For the four
early-onset sets with $L\ge50\mev$ the collapse is a factor of $9.5$, $9.8$,
$11.0$ and $19.0$, with $L=30\mev$ an outlier at $284$ because the hadronic
matter just below its transition is unusually soft and the product starts from
$0.0067$ rather than $0.023$. For the two $n_t=0.5\fm$ sets it is only $2.1$
and $3.2$. That difference is not
incidental: those are precisely the two models in
which the trend of $f_{g_1}$ with mass fails to reverse. The size of the
buoyancy collapse and the sign of $df_{g_1}/dM$ are the same statement seen
twice.

The contrast with the first-order case is worth drawing carefully, because the
two mechanisms are very nearly exact opposites. At a Gibbs mixed phase $\ce^2$ \emph{drops} while
$\cs^2$ stays put; the denominator of Eq.~(\ref{e:ratio}) collapses, the
buoyancy spikes, and $f_{g_1}$ jumps upward. Here $\ce^2$ rises, the
denominator grows, the buoyancy collapses, and $f_{g_1}$ falls. Both are
signatures of quark matter and they point in opposite directions, which is
useful, because it means the two cases cannot be confused with each other.

Behind the arithmetic is the buffering described at the end of
Sec.~\ref{s:quad}. Compress quarkyonic matter at fixed lepton fractions and the
quark densities move immediately, through Eq.~(\ref{e:strong}), to keep
$\mu_d$ and $\mu_u$ consistent with the new $\mu_n$ and $\mu_p$. In doing so
they soak up much of the change in $\mu_n-\mu_p$ that would otherwise appear as
a chemical imbalance, so $\bm A$ in Eq.~(\ref{e:quad}) is small; and the same
shell kinematics that stiffens the equation of state makes $\mu_n$ large, so
the prefactor $\nB^2/\mu_n$ is small as well. Weak driving against a stiff response: no buoyancy. To an excellent
approximation a quarkyonic core is nearly buoyancy-free, and the $g$ mode is
expelled from it into the nucleonic shell outside. Every result in the
remainder of this paper is a consequence of that one sentence.

A small buoyancy survives, and it is worth asking which channel carries it,
because the answer bears directly on what superfluidity would do to these
modes. Equation~(\ref{e:quad}) splits cleanly into three pieces,
\begin{equation}
  \cs^2-\ce^2=\frac{\nB^2}{\mu_n}\Big[A_e^2(\bm M^{-1})_{ee}
  +2A_eA_\mu(\bm M^{-1})_{e\mu}+A_\mu^2(\bm M^{-1})_{\mu\mu}\Big],
  \label{e:chan}
\end{equation}
an electron channel, a muon channel, and the interference between them. We have
evaluated the three separately; at every density in Table~\ref{t:chan} their
sum reproduces the directly differenced $\cs^2-\ce^2$ to six digits, which is
one more independent check on the formalism. The one place the check fails is
a sample taken exactly at a transition density, where the finite difference
straddles the kink; no such point is tabulated. Table~\ref{t:chan} collects the result, and the two families could
hardly behave more differently.

\begin{table}[t]
\caption{Decomposition of $\cs^2-\ce^2$ into lepton channels,
Eq.~(\ref{e:chan}), as a percentage of the total; $e$, $e\mu$ and $\mu$ label
the electron, interference and muon terms. The last column is the ratio of the
$npe$ value $\nB^2A_e^2/(\mu_n M_{ee})$, obtained by freezing the muon
fraction as well, to the full result.}
\label{t:chan}
\begin{ruledtabular}
\begin{tabular}{lccccc}
model & $\nB$ (fm$^{-3}$) & $e$ (\%) & $e\mu$ (\%) & $\mu$ (\%) & $npe$/full \\
\hline
QY $L{=}50$  & 0.35 &  92.8 & $-19.5$ &  26.7 & 0.893 \\
             & 0.50 &  92.3 & $-8.7$  &  16.4 & 0.911 \\
             & 0.70 &  92.6 & $-6.1$  &  13.5 & 0.919 \\
             & 0.90 &  92.8 & $-5.0$  &  12.2 & 0.923 \\
HAD $L{=}50$ & 0.35 &  43.2 & $-33.2$ &  90.0 & 0.402 \\
             & 0.50 &  59.7 & $-44.7$ &  84.9 & 0.539 \\
             & 0.70 &  75.2 & $-61.6$ &  86.5 & 0.642 \\
             & 0.90 &  89.9 & $-86.0$ &  96.2 & 0.706 \\
QY $L{=}70$  & 0.50 &  84.9 & $-13.1$ &  28.2 & 0.834 \\
             & 0.90 &  85.4 & $-7.5$  &  22.1 & 0.848 \\
HAD $L{=}70$ & 0.50 &  98.0 & $-104.3$& 106.3 & 0.724 \\
             & 0.90 & 183.2 & $-273.5$& 190.3 & 0.849 \\
\end{tabular}
\end{ruledtabular}
\end{table}

In the hadronic controls the buoyancy is a delicate residue. The electron and
muon channels are individually enormous and they very nearly cancel: at
$\nB=0.9\fm$ the $L=70\mev$ control has an electron term of $183\%$ and a muon
term of $190\%$ of the total, with an interference term of $-274\%$ undoing
almost all of it. Freezing the muon fraction as well, the $npe$ approximation, captures only $40$--$85\%$ of the answer. The hadronic $g$
mode is supported by a near-cancellation between two lepton channels, and it is
correspondingly sensitive to anything that disturbs either one.

Above the quarkyonic transition the picture is entirely different. One channel
dominates: at the densities sampled in Table~\ref{t:chan} the electron term
carries $85$--$93\%$ of the total, the muon term $12$--$28\%$, and the
interference is a $5$--$20\%$ correction rather than a cancellation. Individual
points elsewhere on the grid lie outside those ranges, the muon term reaching
$41\%$ just above the transition in the $L=70\mev$ set. The $npe$ approximation now recovers
$83$--$92\%$ of the answer. The quarks, in soaking up the change in
$\mu_n-\mu_p$, have removed the muon channel's ability to compete, and what
buoyancy remains rides almost entirely on the electron: equivalently, on the
proton fraction, to which the electron is tied by charge neutrality.

This has a consequence we can state but not yet compute. Andersson and
Comer~\cite{Andersson01} showed that superfluidity removes the proton-fraction
buoyancy, and Kantor and Gusakov~\cite{Kantor14} showed that the $g$ modes
return because the $n_\mu/n_e$ gradient takes over the job; leptonic buoyancy
of exactly this kind also drives the compressional modes of two-superfluid
stars~\cite{RauWasserman18}. In a quarkyonic core that rescue channel is the
minority one. If superfluidity were to suppress the electron term while leaving
the muon term intact, Eq.~(\ref{e:chan}) would retain $12$--$27\%$ of an
already suppressed buoyancy, and $f_{g_1}\propto\sqrt{\cs^2-\ce^2}$ would fall
by a further factor of $0.35$--$0.5$. Superfluidity, in other words, looks
likely to deepen the effect we report rather than undo it. Establishing that
properly requires entrainment and a two-fluid treatment, which we have not
attempted.

\subsection{Structure and observational constraints}

\begin{figure*}[t]
\centerline{\includegraphics[width=\textwidth]{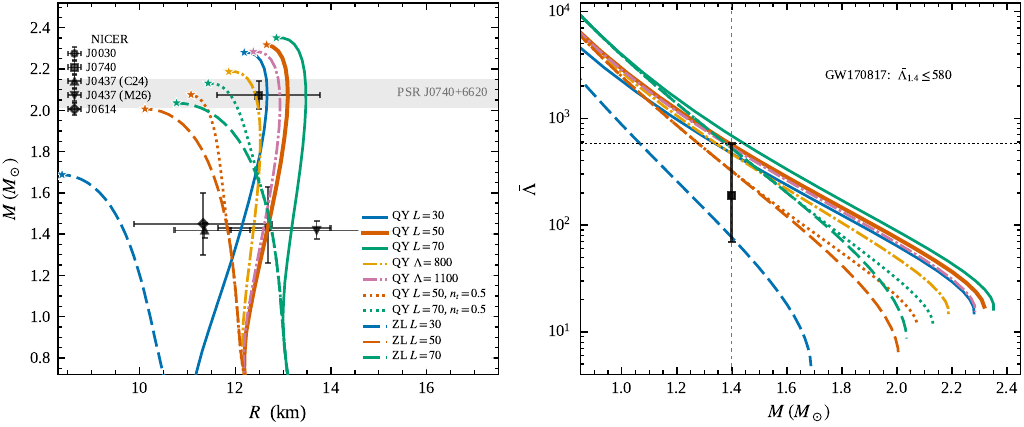}}
\caption{Left: mass--radius relations; stars mark the maximum-mass
configurations. All observational entries are external to this work. The shaded
band is the mass of PSR J0740+6620~\cite{Fonseca21}; the points with error bars
are the NICER pulse-profile inferences for J0030+0451~\cite{Kini26},
J0740+6620~\cite{Salmi24}, J0437$-$4715 (two independent analyses,
C24~\cite{Choudhury24} and M26~\cite{Miller26a}), and
J0614$-$3329~\cite{Miller26b}. Right: dimensionless tidal deformability versus
mass; the square is the GW170817 value $\Lbar_{1.4}=190^{+390}_{-120}$ and the
dotted line the low-spin $90\%$ upper bound
$\Lbar_{1.4}\le580$~\cite{Abbott18}.}
\label{f:mr}
\end{figure*}

Table~\ref{t:struct} and Fig.~\ref{f:mr} collect the global properties. The
quarkyonic branch is stiff: maximum masses run from $2.08$ to $2.35\,\Msun$,
against $1.69$ to $2.04\,\Msun$ for the hadronic controls built on the same
nucleon interaction. Six of the seven quarkyonic sets clear the mass of PSR
J0740+6620, $2.08\pm0.07\,\Msun$~\cite{Fonseca21}, and none of the hadronic
controls does; the seventh, $L=50\mev$ with $n_t=0.5\fm$, falls short at
$2.076\,\Msun$. The
$L=30\mev$ case is the sharpest illustration: the hadronic star reaches
$1.69\,\Msun$ with $R_{1.4}=9.70\km$, while the quarkyonic star built on the
identical low-density interaction reaches $2.28\,\Msun$ with
$R_{1.4}=12.09\km$. This is the stiffening for which quarkyonic matter was
proposed.

Where that mass is carried matters as much as how much of it there is. The
quarkyonic maximum-mass configurations are reached at central densities of
$0.73$--$1.00\fm$, while the hadronic controls need $1.09$--$1.73\fm$ to
reach masses lower by $0.31$, $0.32$ and $0.59\,\Msun$ at $L=50$, $70$ and
$30\mev$ respectively. A quarkyonic star supports more mass at
\emph{less} compression, because the stiffening is spent in the shell just
above $n_t$ rather than in the deep interior. This is the structural
counterpart of the sound-speed profile in Fig.~\ref{f:eos}: the centre of even
the heaviest quarkyonic star sits in the soft, nearly conformal regime beyond
the peak, and that is the region the $g$ mode will find unstratified.

The tidal deformability then works in the opposite direction. $\Lbar_{1.4}$
rises monotonically with $L$ through the quarkyonic sequence, $467$ at
$L=30\mev$, $560$ at $L=50\mev$, $683$ at $L=70\mev$, so the GW170817 low-spin
bound $\Lbar_{1.4}\le580$~\cite{Abbott18} excludes the stiffest symmetry energy
and, interpolating between the $L=50$ and $70\mev$ values, leaves a window
$L\lesssim53\mev$ at $n_t=0.3\fm$. The same conclusion is
visible in the radius: $R_{2.08}=13.48\km$ for $L=70\mev$ sits outside the
$12.35\pm0.75\km$ of Ref.~\cite{Miller21}, whereas the baseline $13.10\km$ is
inside it. Raising $n_t$ to $0.5\fm$ softens the star, the quarkyonic stiffening simply starts later, and pulls $\Lbar_{1.4}$ down to $320$ while
also pulling $M_{\rm max}$ down to $2.08\,\Msun$, at the edge of the mass
constraint.

Table~\ref{t:struct} carries this out model by model. Five of the seven
quarkyonic sets pass every test we can apply; the $L=70\mev$, $n_t=0.3\fm$ set
passes the GW170817 radius test but fails the $\Lbar_{1.4}$ bound, the
Miller \emph{et al.} $R_{1.4}$ range and the $R_{2.08}$ range, and the
$L=50\mev$, $n_t=0.5\fm$ set falls just short of the J0740+6620 mass. None of
the three hadronic controls clears $2.08\,\Msun$ and only one clears
$2.01\,\Msun$. The NICER radius inferences are not at present mutually consistent: the two published analyses of PSR
J0437$-$4715 give $R=11.36^{+0.95}_{-0.63}\km$~\cite{Choudhury24} and a
symmetric $68\%$ range of $11.9$--$15.5\km$~\cite{Miller26a} at essentially the
same mass. We therefore do not use J0437 as a discriminator, and note that our
surviving models sit inside the union of the two.

\begin{table}[t]
\caption{The ten equations of state. Every set shares the same
isoscalar sector, $K_0=220\mev$ and $S_v=31\mev$; only the symmetry-energy
slope $L$ and the two quarkyonic parameters are varied. $\Lambda$ is the
shell-transparency scale and $n_t$ the transition density; a dash marks the
purely nucleonic Zhao--Lattimer controls. The combination
$\eta=(K_0L^2)^{1/3}$ organises the nucleonic $f$-, $p_1$- and $w_1$-mode
relations of Ref.~\cite{Sotani21}, while $\eta_0=L/K_0$ is the combination
entering the nucleonic $g$-mode relation of Ref.~\cite{SotaniTogashi26}; at
fixed $K_0$ both are functions of $L$ alone here, and they are listed for
comparison with that literature. The $L=30\mev$ sets lie outside both fiducial
windows: below the $70.5\lesssim\eta\lesssim114.8\mev$ of Ref.~\cite{Sotani21},
and below the $0.141\le\eta_0\le0.389$ spanned by the equations of state of
Ref.~\cite{SotaniTogashi26}.}
\label{t:eos}
\begin{ruledtabular}
\begin{tabular}{lccccc}
 & $L$ & $\eta$ & $\eta_0$ & $\Lambda$ & $n_t$ \\
Model & (MeV) & (MeV) & & (MeV) & (fm$^{-3}$) \\
\hline
QY$_{30}$ & 30 & 58.3 & 0.136 & $1400$ & 0.3 \\
QY$_{50}$ & 50 & 81.9 & 0.227 & $1400$ & 0.3 \\
QY$_{70}$ & 70 & 102.5 & 0.318 & $1400$ & 0.3 \\
QY$_{50}^{800}$ & 50 & 81.9 & 0.227 & 800 & 0.3 \\
QY$_{50}^{1100}$ & 50 & 81.9 & 0.227 & 1100 & 0.3 \\
QY$_{50}^{\ast}$ & 50 & 81.9 & 0.227 & $1400$ & 0.5 \\
QY$_{70}^{\ast}$ & 70 & 102.5 & 0.318 & $1400$ & 0.5 \\
\hline
ZL$_{30}$ & 30 & 58.3 & 0.136 & -- & -- \\
ZL$_{50}$ & 50 & 81.9 & 0.227 & -- & -- \\
ZL$_{70}$ & 70 & 102.5 & 0.318 & -- & -- \\
\end{tabular}
\end{ruledtabular}
\end{table}

\begin{table*}[t]
\caption{Global properties of the ten sequences, and their standing against
the current constraint set. Model labels are those of Table~\ref{t:eos}.
All bounds are external to this work:
(a) $M_{\rm max}\ge2.01\,\Msun$ (PSR J0348+0432~\cite{Antoniadis13});
(b) $M_{\rm max}\ge2.08\,\Msun$ (PSR J0740+6620~\cite{Fonseca21});
(c) $R_{1.4}=11.9\pm1.4\km$~\cite{Abbott18};
(d) $R_{1.4}=12.45\pm0.65\km$~\cite{Miller21};
(e) $\Lbar_{1.4}\le580$~\cite{Abbott18};
(f) $R_{2.08}=12.35\pm0.75\km$~\cite{Miller21}.
A dash marks a mass the model never reaches, and hence a test it cannot be
subjected to.}
\label{t:struct}
\begin{ruledtabular}
\begin{tabular}{lccc|ccc|c|cccccc}
 & \multicolumn{3}{c|}{maximum mass} & \multicolumn{3}{c|}{$M=1.4\,\Msun$} & $M=2.08\,\Msun$ & \multicolumn{6}{c}{constraint satisfied} \\
Model & $M_{\rm max}$ & $R$ & $n_c$ & $R_{1.4}$ & $\Lbar_{1.4}$ & $n_c$ & $R_{2.08}$ & (a) & (b) & (c) & (d) & (e) & (f) \\
 & ($\Msun$) & (km) & (fm$^{-3}$) & (km) & & (fm$^{-3}$) & (km) & & & & & & \\
\hline
QY$_{30}$ & 2.280 & 12.19 & 0.813 & 12.09 & 466.7 & 0.358 & 12.67 & $\checkmark$ & $\checkmark$ & $\checkmark$ & $\checkmark$ & $\checkmark$ & $\checkmark$ \\
QY$_{50}$ & 2.317 & 12.65 & 0.732 & 12.65 & 560.3 & 0.342 & 13.10 & $\checkmark$ & $\checkmark$ & $\checkmark$ & $\checkmark$ & $\checkmark$ & $\checkmark$ \\
QY$_{70}$ & 2.351 & 12.86 & 0.763 & 13.16 & 683.2 & 0.330 & 13.48 & $\checkmark$ & $\checkmark$ & $\checkmark$ & $\times$ & $\times$ & $\times$ \\
QY$_{50}^{800}$ & 2.187 & 11.86 & 0.901 & 12.37 & 465.8 & 0.395 & 12.42 & $\checkmark$ & $\checkmark$ & $\checkmark$ & $\checkmark$ & $\checkmark$ & $\checkmark$ \\
QY$_{50}^{1100}$ & 2.284 & 12.38 & 0.809 & 12.57 & 531.7 & 0.356 & 12.93 & $\checkmark$ & $\checkmark$ & $\checkmark$ & $\checkmark$ & $\checkmark$ & $\checkmark$ \\
QY$_{50}^{\ast}$ & 2.076 & 11.08 & 0.980 & 11.86 & 320.2 & 0.518 & -- & $\checkmark$ & $\times$ & $\checkmark$ & $\checkmark$ & $\checkmark$ & -- \\
QY$_{70}^{\ast}$ & 2.131 & 11.43 & 0.997 & 12.77 & 520.4 & 0.444 & 11.88 & $\checkmark$ & $\checkmark$ & $\checkmark$ & $\checkmark$ & $\checkmark$ & $\checkmark$ \\
\hline
ZL$_{30}$ & 1.688 & 8.38 & 1.734 & 9.70 & 75.8 & 0.966 & -- & $\times$ & $\times$ & $\times$ & $\times$ & $\checkmark$ & -- \\
ZL$_{50}$ & 2.006 & 10.11 & 1.210 & 11.86 & 322.9 & 0.531 & -- & $\times$ & $\times$ & $\checkmark$ & $\checkmark$ & $\checkmark$ & -- \\
ZL$_{70}$ & 2.035 & 10.77 & 1.094 & 12.77 & 520.7 & 0.444 & -- & $\checkmark$ & $\times$ & $\checkmark$ & $\checkmark$ & $\checkmark$ & -- \\
\end{tabular}
\end{ruledtabular}
\end{table*}

\subsection{Interior profiles}

\begin{figure*}[t]
\centerline{\includegraphics[width=\textwidth]{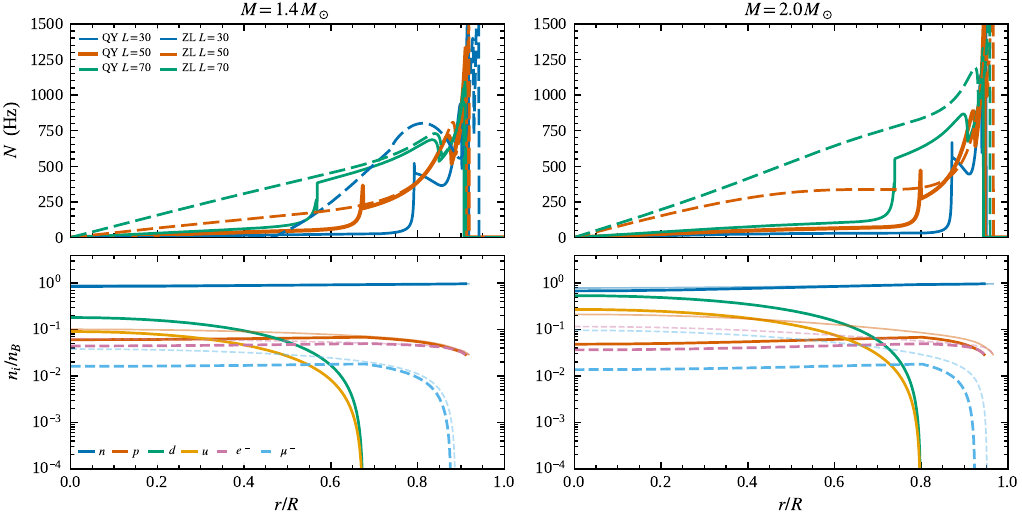}}
\caption{Inside the star. Top: the \bv\ frequency at $1.4$ (left), and
$2.0\,\Msun$ (right); solid quarkyonic, dashed the hadronic controls. Bottom:
the particle fractions through the same stars, heavy curves for the baseline
quarkyonic model and faint curves for its hadronic control, with the same
species colours as Fig.~\ref{f:comp}. The two rows line up: the quark
fractions vanish at $r/R=0.68$ and $0.80$ respectively, and those are exactly
the radii at which $N$ switches on. A narrow spike in $N$ at the
crust--core join, an artefact of the SLy splice discussed in
Sec.~\ref{s:disc}, runs off the top of the panel.}
\label{f:internal}
\end{figure*}

Figure~\ref{f:internal} puts the two halves of the argument on the same axes.
In the hadronic controls $N$ grows smoothly outwards from zero at the centre
and peaks in the outer core, which is the textbook profile for a star whose
composition is stratified all the way down. In the quarkyonic stars $N$ is
pinned near zero through the whole quarkyonic core and then switches on
abruptly. The bottom row says why: the switch-on radius is the radius at which
the quark fractions go to zero. Inside it the composition is quark-dominated
and nearly unstratified; outside it the matter is ordinary $npe\mu$ and
buoyant. The buoyant region is a shell, and comparing the two columns shows
that the shell is thinner in the heavier star, because the transition radius
has moved outward from $r/R=0.68$ to $0.80$.

The figure also closes the argument quantitatively. At $1.4\,\Msun$ the
largest $N$ anywhere inside the quarkyonic core is $365$~Hz, against
$1334$~Hz in the nucleonic shell outside it: a factor of $3.7$ in $N$, and
therefore of $13$ in $N^2$. That is the same factor by which
$\ce^{-2}-\cs^{-2}$ collapses at the transition in Fig.~\ref{f:cs}, computed
there from the equation of state alone with no star attached. The
suppression seen in the stellar profile is neither a structural accident nor a
numerical one; it is the microphysics of Eq.~(\ref{e:quad}) propagated
through the Tolman--Oppenheimer--Volkoff equations without change. Note also
that $N$ vanishes at the centre of \emph{every} model, quarkyonic or not,
because $N^2$ carries a factor $g^2$ and gravity goes to zero
there; what distinguishes the quarkyonic stars is that $N$ then stays small
out to $r_t$ instead of climbing.

Everything in the following subsections is a consequence of that picture.

\subsection{Mode frequencies}

\begin{figure*}[t]
\centerline{\includegraphics[width=\textwidth]{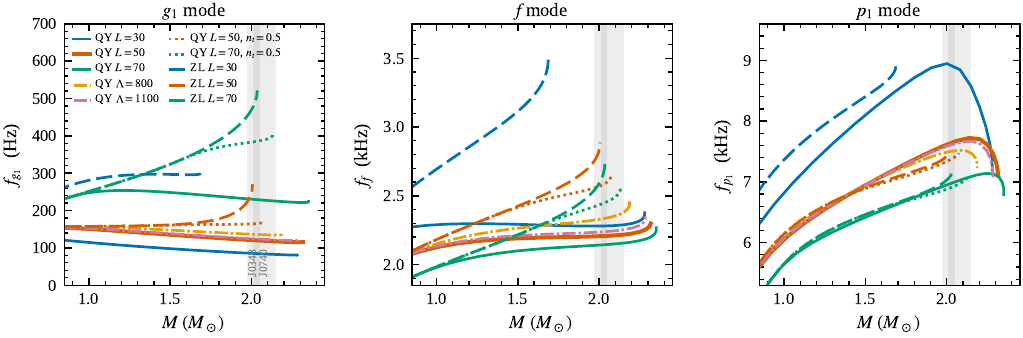}}
\caption{Frequencies of the $g_1$, $f$ and $p_1$ modes versus stellar mass.
Solid curves are quarkyonic with $n_t=0.3\fm$, dotted with $n_t=0.5\fm$, dashed
the hadronic ZL controls; colour encodes the symmetry-energy slope $L$ except
for the two curves that vary $\Lambda$ at fixed $L=50\mev$. All frequencies are
relativistic Cowling values. The vertical bands are the radio-timing masses of
PSR J0740+6620 and PSR J0348+0432~\cite{Fonseca21,Antoniadis13}: any star whose
$g$ mode we might hope to observe has a mass somewhere on this axis, and the
two best-measured ones sit where the quarkyonic and hadronic curves are
furthest apart.}
\label{f:fm}
\end{figure*}

If the buoyancy has been expelled from the core, then two things ought to
follow. The mode should be slower, since it is now supported by a shell rather
than by the whole star. And it should grow slower still as the star gets
heavier, since a heavier star has a larger core and a thinner shell. Both
predictions can be checked, and Fig.~\ref{f:fm} and Table~\ref{t:modes} show
that both are borne out.

Quarkyonic matter \emph{suppresses} the $g$ mode. At $1.4\,\Msun$ the
baseline quarkyonic star has $f_{g_1}=138.4$~Hz against $160.1$~Hz for the
hadronic control with the same $L$; at $2.0\,\Msun$ the gap is much wider,
$121.1$~Hz against $250.7$~Hz. (Table~\ref{t:modes} evaluates the spectrum at
the sequence point nearest the quoted mass, Table~\ref{t:fix} interpolates to
exactly that mass; the two differ by up to $0.4\%$, as at $2.0\,\Msun$, where
the interpolated control gives $249.7$~Hz.) Over the five $n_t=0.3\fm$ sets the quarkyonic
$g_1$ frequencies span $81$--$254$~Hz along the whole stable branch, the
hadronic ones $130$--$522$~Hz. The hadronic lower end belongs to a
sub-solar-mass configuration; restricting the comparison to $M\ge1.0\,\Msun$
leaves the quarkyonic span unchanged at $81$--$254$~Hz and raises the hadronic
one to $158$--$522$~Hz. The quarkyonic minimum is not a light star but the
heaviest one, $80.6$~Hz at $2.28\,\Msun$, which is the reversal itself. The suppression is present in every quarkyonic
model, including those with $n_t=0.5\fm$: at $2.0\,\Msun$ these give $164.2$
and $383.3$~Hz against $250.7$ and $471.4$~Hz for the corresponding controls.
The higher-order $g$ modes behave the same way.

The \emph{trend with mass reverses} once the
quarkyonic core is large enough. In the hadronic stars $f_{g_1}$ increases with
$M$ ($158\to271$~Hz for $L=50\mev$ and $278\to522$~Hz for $L=70\mev$ between $1.2\,\Msun$ and the maximum mass), because the star becomes more compact and
the local gravity $g$ grows. In every $n_t=0.3\fm$ quarkyonic model $f_{g_1}$
\emph{decreases} instead, monotonically from $1.2\,\Msun$ to within
$0.03\,\Msun$ of the maximum mass, where the branch turns over and the last
two or three grid points flatten: $104\to81$~Hz for $L=30\mev$,
$143\to115$~Hz for the baseline, $254\to222$~Hz for $L=70\mev$,
$153\to134$~Hz for $\Lambda=800\mev$ and $147\to120$~Hz for
$\Lambda=1100\mev$. The controls do the opposite without exception: not one step of the $L=50$ or
$L=70\mev$ hadronic sequences decreases.

The reversal emerges from a competition between two effects pulling in
opposite directions, and it is worth separating them. Making a star heavier makes it more compact, which increases the local
gravity and therefore the buoyancy frequency $N\propto g$; this is why the
frequency rises with mass in an ordinary neutron star, and it operates in the
quarkyonic star too. But making a quarkyonic star heavier also pushes the
transition radius outward, converting buoyant nucleonic material into
weakly stratified core and leaving the mode a thinner shell to live in. The
first effect raises the frequency, the second lowers it, and which one wins is
a question about geometry. For $n_t=0.3\fm$ the shell thins fast enough that
the second wins from $1.2\,\Msun$ upward; for $n_t=0.5\fm$ the core is not yet
large enough and the first still wins. Section~\ref{s:shell} makes the
geometrical statement quantitative.

The $n_t=0.5\fm$ models delimit the effect. Their cores reach $n_t$ only at
$1.35\,\Msun$ for $L=50\mev$ and $1.58\,\Msun$ for $L=70\mev$; below
$1.4\,\Msun$ they are indistinguishable from the hadronic controls to better
than $0.1$~Hz, and above onset the quarkyonic core is still too thin a fraction
of the star to reverse the trend: $f_{g_1}$ is pushed down
relative to the control but continues to rise, $158\to169$~Hz for $L=50\mev$
and $285\to406$~Hz for $L=70\mev$. A negative $df_{g_1}/dM$ therefore requires
not merely the presence of quarkyonic matter but a core occupying an
appreciable fraction of the star, which on this grid means an onset at or below
$\sim0.3\fm$.

The overtone spacing behaves in a way worth recording, though it does
\emph{not} work as a signature and we do not offer it as one. The $L=50$ and
$L=70\mev$ controls give $f_{g_2}/f_{g_1}=0.710\pm0.044$ and
$0.667\pm0.006$, stable along each nucleonic sequence, while the quarkyonic
models with $n_t=0.3\fm$ give $0.49$--$0.64$: the overtones are squeezed into
the nucleonic shell while the fundamental is not. But the $L=30\mev$ control
gives $0.46$--$0.56$, lower than every quarkyonic model, because its own deep
interior is buoyancy-free for the unrelated reason set out in
Sec.~\ref{s:lim}. A low overtone ratio therefore signals a thin buoyant
region, whatever produced it, and cannot distinguish a quarkyonic core from a
soft symmetry energy. The overtones are in any case the modes most vulnerable
to the finite reaction rates discussed in Sec.~\ref{s:disc}.

At the two reference masses the $f$-mode frequencies cluster at
$2.1$--$3.0$~kHz and the $p_1$ at $6.3$--$8.9$~kHz (over the whole stable
branch above $1.0\,\Msun$, $2.0$--$3.5$ and $5.8$--$9.0$~kHz), with quarkyonic
and hadronic values separated by less than the spread among nuclear
parameters. As in Ref.~\cite{Zheng23}, neither is a
useful discriminator on its own.

The cleanest way to see all of this is to hold two things fixed at once, the mass and the nuclear interaction, and change nothing but the presence of the
quarks. Table~\ref{t:fix} does that, comparing each quarkyonic model with the
purely hadronic control built on the identical low-density interaction, star by
star along a common mass grid. Three patterns stand out. The $g_1$ ratio falls
steadily with mass, from $0.91$ at $1.2\,\Msun$ to $0.49$ at $2.0\,\Msun$ for
both $L=50$ and $L=70\mev$: the same number for two different symmetry
energies, which is what one expects if the suppression is geometrical rather
than compositional in origin. The $p_1$ ratio, in contrast, sits at unity to
within $2.7\%$ at every mass, so at fixed mass the $p_1$ mode carries no
information about the transition whatsoever. And the $f$ ratio departs from
unity by as much as $24\%$, not because the $f$ mode is sensitive to the quarks
but because the two stars have different radii at the same mass; that is
exactly why the $f$ mode must be compared at fixed compactness, as we do in
Sec.~\ref{s:univ}.

\begin{table}[t]
\caption{The two families compared at fixed mass and fixed symmetry-energy
slope. QY denotes the quarkyonic model with $\Lambda=1400\mev$ and
$n_t=0.3\fm$, HAD the purely hadronic control built on the same nucleon
interaction; both are interpolated onto a common mass grid along the stable
branch. Radii are in km and frequencies in Hz. The $L=30\mev$ control reaches
only $1.69\,\Msun$, so no entries exist for it above $1.6\,\Msun$.}
\label{t:fix}
\begin{ruledtabular}
\begin{tabular}{lcccrrccc}
 & $M$ & $R_{\rm QY}$ & $R_{\rm HAD}$ & $f_{g_1}^{\rm QY}$
 & $f_{g_1}^{\rm HAD}$ & \multicolumn{3}{c}{QY/HAD ratio} \\
\cline{7-9}
 & ($\Msun$) & & & & & $g_1$ & $f$ & $p_1$ \\
\hline
\multirow{3}{*}{$L{=}30$}  & 1.2 & 11.82 & 10.00 &  107.4 &  293.8 & 0.366 & 0.805 & 0.941 \\
                           & 1.4 & 12.09 &  9.71 &  101.0 &  297.2 & 0.340 & 0.760 & 0.957 \\
                           & 1.6 & 12.33 &  9.17 &   95.2 &  296.1 & 0.322 & 0.709 & 0.968 \\
\multirow{5}{*}{$L{=}50$}  & 1.2 & 12.46 & 11.97 &  144.8 &  158.4 & 0.914 & 0.948 & 0.989 \\
                           & 1.4 & 12.65 & 11.85 &  138.5 &  160.1 & 0.865 & 0.919 & 1.000 \\
                           & 1.6 & 12.83 & 11.66 &  132.3 &  164.7 & 0.803 & 0.887 & 1.013 \\
                           & 1.8 & 12.98 & 11.33 &  126.4 &  177.9 & 0.711 & 0.850 & 1.027 \\
                           & 2.0 & 13.08 & 10.35 &  121.1 &  249.7 & 0.485 & 0.779 & 1.024 \\
\multirow{5}{*}{$L{=}70$}  & 1.2 & 13.05 & 12.90 &  253.7 &  278.0 & 0.913 & 0.984 & 0.992 \\
                           & 1.4 & 13.16 & 12.78 &  251.2 &  307.5 & 0.817 & 0.961 & 0.991 \\
                           & 1.6 & 13.29 & 12.57 &  245.2 &  342.5 & 0.716 & 0.931 & 0.995 \\
                           & 1.8 & 13.40 & 12.22 &  237.9 &  388.3 & 0.613 & 0.893 & 0.997 \\
                           & 2.0 & 13.47 & 11.38 &  230.5 &  470.4 & 0.490 & 0.827 & 0.988 \\
\end{tabular}
\end{ruledtabular}
\end{table}

\begin{table*}[t]
\caption{Quadrupole ($l=2$) mode frequencies at $M=1.4$ and $2.0\,\Msun$, with the $g_1$-mode gravitational-wave damping time and strain amplitude for a mode energy $E=10^{51}$~erg. All frequencies are relativistic Cowling values and are therefore lower bounds (Sec.~\ref{s:num}). The $g_3$ mode is computed but not listed: finite weak-reaction rates are expected to remove the high overtones~\cite{Andersson19,Counsell24}, and we use it only for the node-count check.}
\label{t:modes}
\begin{ruledtabular}
\begin{tabular}{lcrrrrrccc}
 & $M$ & \multicolumn{5}{c}{$l=2$ frequencies (Hz)} & $\tau_{\rm GW}$ & \multicolumn{2}{c}{$|h_+|$} \\
\cline{3-7}\cline{9-10}
Model & ($M_\odot$) & $g_2$ & $g_1$ & $f$ & $p_1$ & $p_2$ & (s) & $10$ kpc & $15$ Mpc \\
\hline
\multirow{2}{*}{QY$_{30}$} & 1.4 & 60.9 & 101.0 & 2296 & 7870 & 9864 & $1.2\times10^{15}$ & $1.3\times10^{-26}$ & $8.5\times10^{-30}$ \\
 & 2.0 & 50.3 & 85.6 & 2280 & 8946 & 10686 & $3.7\times10^{12}$ & $2.7\times10^{-25}$ & $1.8\times10^{-28}$ \\
\multirow{2}{*}{QY$_{50}$} & 1.4 & 77.2 & 138.4 & 2185 & 6794 & 8824 & $5.3\times10^{10}$ & $1.4\times10^{-24}$ & $9.3\times10^{-28}$ \\
 & 2.0 & 65.8 & 121.1 & 2206 & 7614 & 10721 & $3.4\times10^{13}$ & $6.3\times10^{-26}$ & $4.2\times10^{-29}$ \\
\multirow{2}{*}{QY$_{70}$} & 1.4 & 138.4 & 251.3 & 2086 & 6302 & 8137 & $4.8\times10^{8}$ & $8.1\times10^{-24}$ & $5.4\times10^{-27}$ \\
 & 2.0 & 115.3 & 230.4 & 2142 & 6951 & 10342 & $1.5\times10^{10}$ & $1.6\times10^{-24}$ & $1.0\times10^{-27}$ \\
\multirow{2}{*}{QY$_{50}^{800}$} & 1.4 & 92.7 & 150.0 & 2249 & 6808 & 9087 & $4.9\times10^{10}$ & $1.3\times10^{-24}$ & $8.9\times10^{-28}$ \\
 & 2.0 & 81.2 & 136.5 & 2329 & 7492 & 10401 & $3.7\times10^{11}$ & $5.4\times10^{-25}$ & $3.6\times10^{-28}$ \\
\multirow{2}{*}{QY$_{50}^{1100}$} & 1.4 & 81.1 & 141.4 & 2203 & 6800 & 8905 & $5.1\times10^{10}$ & $1.4\times10^{-24}$ & $9.3\times10^{-28}$ \\
 & 2.0 & 69.8 & 124.7 & 2236 & 7586 & 10673 & $5.1\times10^{12}$ & $1.6\times10^{-25}$ & $1.1\times10^{-28}$ \\
\multirow{2}{*}{QY$_{50}^{\ast}$} & 1.4 & 109.0 & 160.1 & 2376 & 6795 & 9432 & $3.5\times10^{10}$ & $1.5\times10^{-24}$ & $9.8\times10^{-28}$ \\
 & 2.0 & 113.8 & 164.2 & 2569 & 7336 & 11157 & $1.4\times10^{10}$ & $2.2\times10^{-24}$ & $1.5\times10^{-27}$ \\
\multirow{2}{*}{QY$_{70}^{\ast}$} & 1.4 & 206.0 & 307.5 & 2170 & 6356 & 8452 & $2.4\times10^{8}$ & $9.4\times10^{-24}$ & $6.2\times10^{-27}$ \\
 & 2.0 & 228.5 & 383.3 & 2434 & 6880 & 10493 & $1.5\times10^{7}$ & $2.9\times10^{-23}$ & $2.0\times10^{-26}$ \\
\hline
\multirow{1}{*}{ZL$_{30}$} & 1.4 & 149.1 & 297.2 & 3021 & 8228 & 11922 & $9.4\times10^{9}$ & $1.5\times10^{-24}$ & $1.0\times10^{-27}$ \\
\multirow{2}{*}{ZL$_{50}$} & 1.4 & 108.8 & 160.1 & 2376 & 6795 & 9427 & $3.4\times10^{10}$ & $1.5\times10^{-24}$ & $1.0\times10^{-27}$ \\
 & 2.0 & 180.5 & 250.7 & 2835 & 7438 & 10686 & $2.7\times10^{8}$ & $1.1\times10^{-23}$ & $7.3\times10^{-27}$ \\
\multirow{2}{*}{ZL$_{70}$} & 1.4 & 206.0 & 307.5 & 2170 & 6356 & 8454 & $2.4\times10^{8}$ & $9.4\times10^{-24}$ & $6.3\times10^{-27}$ \\
 & 2.0 & 313.5 & 471.4 & 2593 & 7033 & 10197 & $2.6\times10^{6}$ & $5.9\times10^{-23}$ & $3.9\times10^{-26}$ \\
\end{tabular}
\end{ruledtabular}
\end{table*}

\subsection{Why the trend reverses}
\label{s:shell}

\begin{figure*}[t]
\centerline{\includegraphics[width=0.86\textwidth]{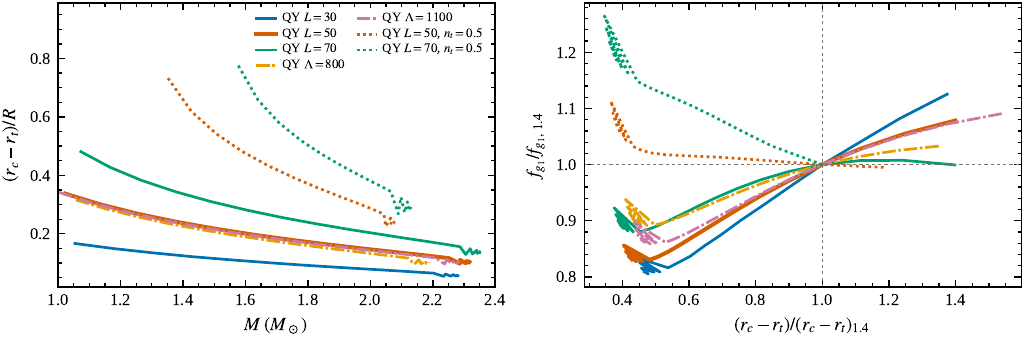}}
\caption{Left: the fractional thickness of the buoyant nucleonic shell,
$(r_c-r_t)/R$, where $r_t$ is the radius at which $n_B=n_t$ and $r_c$ the
crust--core boundary. Right: the $g_1$ frequency and the shell thickness, each
normalised to its value at $1.4\,\Msun$, for the same models.}
\label{f:shell}
\end{figure*}

The mechanism argued for above, a weakly stratified quarkyonic core with the
mode confined to the nucleonic shell outside it, makes a quantitative
prediction, and Fig.~\ref{f:shell} tests it. As the mass
rises the transition radius moves outwards and the shell thins: for the
baseline, $(r_c-r_t)/R$ falls from $0.35$ to $0.10$ between $1.0\,\Msun$ and
the maximum mass, and the enclosed core grows from $30\%$ to $83\%$ of the
stellar mass. Normalising both quantities to their values at $1.4\,\Msun$, the
$g_1$ frequency and the shell thickness track one another closely in every
$n_t=0.3\fm$ model, with linear correlation coefficients between $0.94$ and
$1.00$. They also very nearly collapse: a single straight line through all
five models leaves a $3\%$ rms and a $7\%$ maximum residual. The shell
geometry is not merely correlated with the trend, it accounts for it. In the
$n_t=0.5\fm$ models the correlation is negative, because there the shell is
still thick. That is a correlation, not yet an explanation, and the rest of
this subsection replaces it with an identity.

\label{s:decomp}

The shell argument is geometrical and qualitative. It can be made exact, and
doing so separates cleanly what the three modes are telling us, and shows
that only one of them is telling us about the quarks at all.

Every stellar oscillation frequency carries the dynamical frequency of the star
as an overall scale. Write
\begin{equation}
  \fdyn \equiv \frac{1}{2\pi}\sqrt{\frac{GM}{R^3}},
  \qquad
  f = \Big(\frac{f}{\fdyn}\Big)\,\fdyn ,
  \label{e:fdyn}
\end{equation}
and take the logarithmic mass derivative of the product. The result is an
identity,
\begin{equation}
  \frac{d\ln f}{d\ln M} = \underbrace{\frac{d\ln (f/\fdyn)}{d\ln M}}_{\Scomp}
  \;+\;\underbrace{\frac{1}{2}\Big(1-3\frac{d\ln R}{d\ln M}\Big)}_{\Sst},
  \label{e:decomp}
\end{equation}
with no approximation anywhere. The second term is fixed entirely by the
mass--radius relation: it is the same for all three modes of a given star and
knows nothing about oscillations. The first is the slope of a dimensionless
number, the mode frequency in units of the star's own dynamical frequency, and it is the only place where the internal physics can enter. Table
\ref{t:decomp} evaluates both terms as chord averages between $1.2$ and
$2.0\,\Msun$; the two columns sum to the third to better than $1.4\times10^{-3}$,
which is the interpolation error and confirms that we have not smuggled in an
approximation.

\begin{table*}[t]
\caption{Decomposition of the mass trend of each mode into a structural and a
compositional part, Eq.~(\ref{e:decomp}), as chord-averaged logarithmic slopes
between $1.2$ and $2.0\,\Msun$. $\Sst=\frac12(1-3\,d\ln R/d\ln M)$ is the slope
of the dynamical frequency $\fdyn=(2\pi)^{-1}(GM/R^3)^{1/2}$ and depends only
on the mass--radius relation; $\Scomp$ is the slope of the dimensionless ratio
$f/\fdyn$, and is the only term into which the internal physics enters. Their
sum, listed as ``total'', is the observable slope $d\ln f/d\ln M$. ZL
$L{=}30\mev$ is omitted because its stable branch ends at $1.69\,\Msun$.}
\label{t:decomp}
\begin{ruledtabular}
\begin{tabular}{lcc|cc|cc|cc}
 & & & \multicolumn{2}{c|}{$g_1$} & \multicolumn{2}{c|}{$f$} & \multicolumn{2}{c}{$p_1$} \\
Model & $d\ln R/d\ln M$ & $\Sst$ & $\Scomp$ & total & $\Scomp$ & total & $\Scomp$ & total \\
\hline
QY $L{=}30$ & $+0.133$ & $+0.301$ & $-0.747$ & $-0.445$ & $-0.317$ & $-0.015$ & $+0.075$ & $+0.377$ \\
QY$_{50}^{800}$ & $+0.035$ & $+0.448$ & $-0.681$ & $-0.232$ & $-0.345$ & $+0.103$ & $-0.170$ & $+0.278$ \\
QY$_{50}^{1100}$ & $+0.082$ & $+0.377$ & $-0.702$ & $-0.324$ & $-0.325$ & $+0.052$ & $-0.067$ & $+0.311$ \\
QY $L{=}50$ & $+0.096$ & $+0.356$ & $-0.707$ & $-0.350$ & $-0.318$ & $+0.038$ & $-0.035$ & $+0.321$ \\
QY $L{=}70$ & $+0.061$ & $+0.408$ & $-0.596$ & $-0.188$ & $-0.317$ & $+0.091$ & $-0.133$ & $+0.276$ \\
QY $L{=}50$, $n_t{=}0.5$ & $-0.091$ & $+0.636$ & $-0.566$ & $+0.071$ & $-0.404$ & $+0.232$ & $-0.410$ & $+0.227$ \\
QY $L{=}70$, $n_t{=}0.5$ & $-0.130$ & $+0.695$ & $-0.066$ & $+0.629$ & $-0.386$ & $+0.310$ & $-0.455$ & $+0.240$ \\
ZL $L{=}50$ & $-0.286$ & $+0.929$ & $-0.039$ & $+0.891$ & $-0.506$ & $+0.423$ & $-0.676$ & $+0.253$ \\
ZL $L{=}70$ & $-0.244$ & $+0.866$ & $+0.163$ & $+1.030$ & $-0.435$ & $+0.431$ & $-0.585$ & $+0.282$ \\
\end{tabular}
\end{ruledtabular}
\end{table*}

Three things follow, and the third is the central result of this paper stated
in its sharpest form.

First, the structural term is itself a quarkyonic signature, and an
unexpected one. A hadronic star contracts as it gains mass: $d\ln R/d\ln M=-0.29$ and
$-0.24$ for the two controls, giving $\Sst=+0.93$ and $+0.87$. An early-onset
quarkyonic star does the opposite (it \emph{expands}, by $1.8$ to $7.0\%$ between $1.2$ and $2.0\,\Msun$), because the stiffening just above $n_t$
holds the outer layers up while mass is added to the core. Its $\Sst$ is
therefore only $+0.30$ to $+0.45$, a factor of two to three smaller. The
late-onset models sit in between at $+0.64$ and $+0.70$, as they must, since
their transition has barely begun. Nothing about oscillations has been used to
get this; it is a restatement of the near-verticality of the quarkyonic
branches in Fig.~\ref{f:mr}.

Second, for the $f$ and $p_1$ modes that structural term is the whole story.
The compositional slopes of the $f$ mode are $-0.32$ to $-0.40$ for every
quarkyonic model and $-0.44$ to $-0.51$ for the controls: the same size and
the same sign, though the matched-$L$ pairs differ by up to $59\%$. The flatness of the quarkyonic
$f$ mode reported in Sec.~\ref{s:obs} is therefore not a statement about the
$f$ mode at all: measured in units of $\fdyn$ the quarkyonic and hadronic $f$
modes behave alike, and the entire difference in the observable slope
($+0.04$ against $+0.42$ for $L=50\mev$) is inherited from $\Sst$, that is,
from the mass--radius relation. The same holds for $p_1$, with the two terms
now partly reinforcing: $\Scomp$ runs from $+0.08$ to $-0.17$ across the
early-onset models (small, and not even of one sign) against $-0.59$ to
$-0.68$ for the controls, and combined with the smaller $\Sst$ it produces the
turnover. These modes are worth measuring, but
what they measure is the mass--radius relation, which NICER already measures
more directly.

Third, and this is the point, the $g_1$ mode is different in kind. For the
nucleonic controls
$\Scomp$ is $-0.04$ and $+0.16$: to within the accuracy of this comparison the
hadronic $g_1$ frequency is simply $\fdyn$ times a constant,
\begin{equation}
  f_{g_1} \simeq k\,\fdyn, \qquad
  k \simeq \begin{cases} 0.095, & L=50\mev,\\ 0.205, & L=70\mev,\end{cases}
  \label{e:kfac}
\end{equation}
quoted at $1.4\,\Msun$ and varying by only $-2\%$ and $+9\%$ between $1.2$
and $2.0\,\Msun$. The
dimensionless number $k$ is the mean buoyancy of the star in units of its
dynamical frequency, and it is set by the composition gradient, which is why
it doubles between $L=50$ and $70\mev$ while $\fdyn$ itself, at fixed mass,
changes by only about a tenth. That $f_{g_1}$ should track $L$ this strongly is
consistent with Ref.~\cite{Sun25}, who find the $g_1$ frequency of nucleonic
stars rising linearly with $L$ across ten relativistic functionals at fixed
mass. For
the early-onset quarkyonic models, by contrast, $\Scomp=-0.60$ to $-0.75$: $k$
is not constant but falls by a quarter to a third between $1.2$ and
$2.0\,\Msun$. Other entries in Table~\ref{t:decomp} are comparable in size, the controls' $p_1$ slopes are $-0.59$ and $-0.68$, but this is the only one
in which the compositional term both dominates its structural partner and
separates the two families. That is what it means to say the $g$
mode measures composition and the others measure structure.

The decomposition also settles, quantitatively, why a late onset fails to
reverse the trend. Both $n_t=0.5\fm$ models have a substantial compositional
suppression once their cores appear ($\Scomp=-0.57$ for $L=50\mev$, within $24\%$ of the early-onset values), but their radii still contract,
so $\Sst=+0.64$ overwhelms it and the observable slope stays positive at
$+0.07$. The sign reversal requires \emph{both} halves of the quarkyonic
effect: the compositional suppression, and the structural stiffening that stops
the star from shrinking. A model that produced one without the other would not
show it. This supersedes the shell-thickness correlation above: the shell thins in the
$n_t=0.5\fm$ models too, and the reason they behave differently is in the
second column of Table~\ref{t:decomp}, not in the geometry.

One consistency check is worth recording, because it ties the decomposition
back to the microphysics. If $g_1$ is a cavity mode trapped by the buoyancy,
then in the WKB limit its frequency should track the cavity integral
$\mathcal{I}_N=\sqrt{l(l+1)}\,(2\pi)^{-1}\int N\,dr/r$. We evaluate this over
the whole star rather than over the buoyant shell alone, because the $g_1$
eigenfunction has amplitude outside the shell; restricting the integral to
$r_t\le r\le r_c$ changes the numbers below but not the trend. Evaluated that
way on the profiles of Fig.~\ref{f:internal}, $\mathcal{I}_N$ reproduces the
sign of the mass trend it is meant to explain: for the baseline quarkyonic
model it falls from $76.1$ to $73.5$~Hz between $1.4$ and $2.0\,\Msun$ while
$f_{g_1}$ falls from $138$ to $121$~Hz, and for its control it rises from
$140$ to $252$~Hz while $f_{g_1}$ rises from $160$ to $251$~Hz. The reversal
is present in the cavity integral itself, before any eigenvalue is computed.
The check is qualitative only, and we do not lean on it: the quarkyonic fall
is $3\%$ against the $12\%$ fall of the eigenfrequency, $n=1$ is not the
asymptotic regime, and the ratio $f_{g_1}/\mathcal{I}_N$ ranges from $1.0$ for
the $L=50\mev$ control to $2.5$ for the $L=30\mev$ one, whose leptons vanish
in the deep interior (Sec.~\ref{s:lim}).

\subsection{Universal relations}
\label{s:univ}

\begin{figure*}[t]
\centerline{\includegraphics[width=\textwidth]{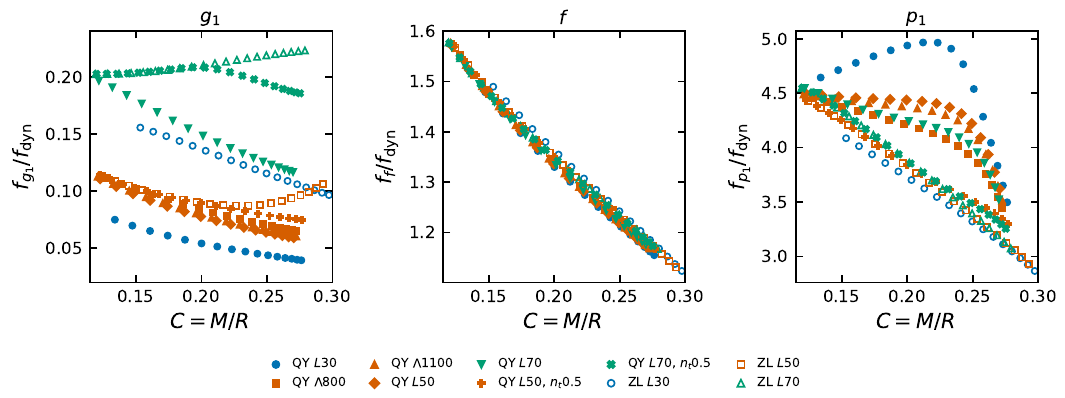}}
\caption{The three modes in the form Eq.~(\ref{e:whyuniv}) says is the
meaningful one: the dimensionless ratio $f/\fdyn$ against compactness, for
every stellar model of all ten equations of state. No fit is drawn and none is
needed: a universal relation exists for a mode if and only if its points
here fall on a single curve. Filled symbols are quarkyonic, open symbols the
nucleonic controls, and colour encodes the symmetry-energy slope $L$, so each
quarkyonic model can be read against the control of the same colour. The $f$
mode collapses to better than $2.5\%$; the $p_1$ mode spreads by a quarter to
a third; the $g_1$ panel does not collapse at all. In it the five early-onset
models lie below the control of their own colour at every compactness, which
is the suppression this paper is about. The two $n_t=0.5\fm$ models instead
track their controls until their cores appear, and the $\Lambda=800\mev$ set
touches its control at the lowest compactness plotted; in all three cases the
ratio reaches $1.00$ and never exceeds it.}
\label{f:univ}
\end{figure*}

Oscillation frequencies are usually reported through relations that are
insensitive to the equation of state, and there is a standard way of doing it:
scale the frequency by a power of the mass, plot it against the compactness
$C=M/R$, fit a low-order polynomial, and show the residual
underneath~\cite{Sotani21,SotaniKumar21,Sotani25,SotaniTogashi26}. We report
the fits in that form, in Table~\ref{t:lam}, so that the numbers can be
compared with the literature. But Sec.~\ref{s:decomp} makes the fitting
unnecessary, and it is worth seeing why before looking at any residual.

Since $\fdyn\,M\propto M^{3/2}R^{-3/2}\propto C^{3/2}$, the conventionally
scaled frequency is
\begin{equation}
  f\,M_{1.4} \;\propto\; \Big(\frac{f}{\fdyn}\Big)\,C^{3/2},
  \qquad M_{1.4}\equiv\frac{M}{1.4\,\Msun},
  \label{e:whyuniv}
\end{equation}
so a relation $fM_{1.4}=P(C)$ exists \emph{if and only if} the dimensionless
ratio $f/\fdyn$ is itself a function of compactness alone. A universal relation
is not an empirical fact to be discovered by fitting; it is the statement that
one particular dimensionless number is fixed by the star's bulk structure. The
$f$ and $p_1$ modes have every reason to satisfy it, because a compressional
mode is a bulk-structure mode. The $g$ mode has none, because its $f/\fdyn$ is
the constant $k$ of Eq.~(\ref{e:kfac}), set by the composition gradient, which
doubles between $L=50$ and $70\mev$ while $\fdyn$ itself changes by about a
tenth.

Figure~\ref{f:univ} therefore plots $f/\fdyn$ directly, with no fit. The test
is the spread at fixed compactness across all ten equations of state, which
involves no arbitrary choice of polynomial order and no split into families:
\begin{equation}
  \frac{\max-\min}{\rm mean}\bigg|_{C}=
  \begin{cases}
    2.2\text{--}2.4\%, & f,\\
    23\text{--}35\%, & p_1,\\
    120\text{--}171\%, & g_1,
  \end{cases}
  \label{e:spread}
\end{equation}
evaluated at $C=0.18$, $0.22$ and $0.26$. Two orders of magnitude separate the
$f$ mode from the $g$ mode. That ordering is not peculiar to our grid: working
in full general relativity with a large library of equations of state, Kuan
\emph{et al.}~\cite{Kuan22} find that pressure-driven modes follow a universal
relation regardless of microphysics while the $g$-mode space splits into groups
labelled by the class of equation of state. Equation~(\ref{e:whyuniv}) is the
reason: a compressional mode has a universal $f/\fdyn$ and a buoyancy-driven
one does not, and the quarkyonic models are a new group in their sense. Note also that $M_{1.4}$ is
\emph{dimensionless}, so the scaled combinations quoted below are frequencies
in kHz and not the kHz$\,\Msun$ that the alternative convention
$f\times(M/\Msun)$ would give.

The $f$ mode measures compactness and nothing else. Fitted as
$f_fM_{1.4}$ against $C$ on every stellar model above $1.0\,\Msun$, which
spans $0.119\le C\le0.297$, the nucleonic controls collapse onto
\begin{equation}
  f_f M_{1.4} = -0.0031 + 8.534\,C + 44.63\,C^2 - 86.59\,C^3\ \ {\rm kHz},
  \label{e:ffit}
\end{equation}
with an rms deviation of $0.40\%$ and a maximum of $1.1\%$, and the quarkyonic
models onto
\begin{equation}
  f_f M_{1.4} = -0.0575 + 9.998\,C + 33.65\,C^2 - 65.24\,C^3\ \ {\rm kHz},
  \label{e:ffitq}
\end{equation}
with $0.43\%$ and $0.82\%$. The two cubics differ by less than the scatter
about either: a single cubic through all ten equations of state holds to
$0.61\%$ rms and $1.4\%$ at worst. Put the comparison at fixed compactness and
the same conclusion is quantitative: a quarkyonic star and the hadronic
control built on the same nuclear interaction differ in $f_fM_{1.4}$ by
$1\%$ at $C=0.265$. The $f$ mode is blind to the transition, which is the
conclusion Ref.~\cite{SotaniKojo23} reached for a quark--hadron crossover.

The $p_1$ mode carries a weak version of the signal, and only at fixed
compactness. The controls follow one cubic to $1.3\%$ rms; the quarkyonic
models scatter about theirs by $6.8\%$ rms and $19\%$ at worst
(Table~\ref{t:lam}), and the scatter is not random; it grows with compactness
and is ordered by $L$. Compared with its own control at the same $C$ and the
same $L$, the ratio of $f_{p_1}M_{1.4}$ rises from unity at low compactness to
$1.19$ ($L=50\mev$), and $1.14$ ($L=70\mev$) at $C=0.265$, with a maximum of
$1.25$ at $C=0.24$ for $L=50\mev$ and $1.41$ there for $L=30\mev$. So the $p_1$
mode does shift, by a fifth and in the softest case by two fifths, but only
once the star is compact enough
for the quarkyonic core to occupy most of it. The caveat is important, and Table~\ref{t:fix} makes it concrete: compared at
fixed \emph{mass} the two families agree on $f_{p_1}$ to within $2.7\%$ at
$L=50$ and $70\mev$, and to within $5.9\%$ at $L=30\mev$. A measurement would therefore have to supply the
compactness independently before any of this became usable.

The $g_1$ mode obeys no relation of this kind at all. Scaled as
$f_{g_1}M_{1.4}$ and fitted the same way, the residuals are $35\%$ rms for the
controls and $47\%$ for the quarkyonic models, reaching $114\%$ across all ten
sets. The failure is not an artefact of the scaling: fitting the unscaled
$f_{g_1}$ leaves $30\%$ rms for the controls, $f_{g_1}M_{1.4}$ leaves $35\%$
and $f_{g_1}M^2$, the scaling used in Ref.~\cite{SotaniTogashi26}, leaves $42\%$, so no power of the mass rescues the relation,
and the tidal-deformability fits below show that replacing $C$ by $\Lbar$ does not either. No
polynomial
in a single structural variable describes the $g$ mode, because the $g$ mode is
not a function of structure alone; it is a function of composition, and two
stars of the same compactness can be composed differently.

It is worth being careful about what this does and does not establish, because
the nucleonic literature does not claim a $g$-mode relation in a structural
variable alone. Sotani and Togashi~\cite{SotaniTogashi26} derive an empirical
relation for $f_{g_1}M^2$ in nucleonic stars as a cubic in the compactness
whose coefficients are themselves cubics in the saturation combination
$\eta_0\equiv L/K_0$: a two-parameter form, accurate to about $10\%$ over
$K_0=223$--$263\mev$ and $L=34.4$--$101\mev$. They propose precisely the test
this paper carries out: a measured $g$-mode frequency departing from their
relation would signal additional degrees of freedom or a new composition inside
the star. Our quarkyonic models make that test fire. At matched compactness and
matched nuclear interaction, hence matched $\eta_0$, the quarkyonic
$f_{g_1}$ is $28$ to $45\%$ below its control for $L=50\mev$ and $41$ to $54\%$
below for $L=70\mev$ across $C=0.22$--$0.265$, three to five times the scatter
of their fit. The earlier relations for the $f$, $p_1$ and $w_1$ modes do not treat $g$ modes
at all: Ref.~\cite{Sotani21} fits them against the compactness and the
different combination $\eta=(K_0L^2)^{1/3}$, and Ref.~\cite{SotaniKumar21}
against the tidal deformability.

Our grid cannot test the two-parameter form itself: $K_0$ and $S_v$ are held
fixed and only $L$ varies, over three values, one of which ($L=30\mev$) loses
its buoyancy entirely in the deep interior. Two clean nucleonic points cannot
constrain a relation in $(C,\eta_0)$. What we have shown is the weaker and
sharper statement that no function of a \emph{structural} variable alone will
do, which is exactly what Sec.~\ref{s:decomp} predicts, since $f_{g_1}/\fdyn$
is set by the composition gradient.

That distinction strengthens rather than weakens the quarkyonic result,
because our internal comparison holds the nuclear interaction fixed as well as
the compactness, and at fixed $K_0$, fixing $L$ fixes $\eta_0$. At fixed compactness and fixed $L$,
$f_{g_1}M_{1.4}$ of the quarkyonic star falls to $0.65$ ($L=50\mev$), and
$0.54$ ($L=70\mev$) of its control by $C=0.265$; at fixed \emph{mass} the suppression
is stronger still, $f_{g_1}$ reaching $0.49$ of the control at $2.0\,\Msun$ for
both. Of the three modes, only the $g$ mode moves by a factor rather than a
few per cent. Whatever function of $(C,\eta)$ the nucleonic $g$ mode may
satisfy, the quarkyonic star with the same $C$ and the same $\eta$ does not
satisfy it, and misses it by a factor of two.

Everything so far is organised by the compactness, and the compactness is not what a
gravitational-wave inspiral measures. It measures the masses and, from the
phase, $\Lbar$; the radius, and hence $C$, follows only through an
equation-of-state assumption, which is precisely the assumption an asteroseismic
test is meant to avoid. We therefore repeat every fit above against
$\log_{10}\Lbar$, on the identical sample and with the identical residual
statistic.

It works exactly as well as $C$. Fitting
$f_fM_{1.4}$, $f_{p_1}M_{1.4}$ and $f_{g_1}M_{1.4}$ to a cubic in
$\log_{10}\Lbar$ over $M\ge1.0\,\Msun$ gives the residuals collected in
Table~\ref{t:lam}, alongside the compactness residuals above
recomputed on the identical sample. No entry changes by more than a factor of
$2.9$, and the ranking of the three modes is untouched. For the nucleonic
controls the $f$-mode relation is
\begin{equation}
  f_f M_{1.4} = 4.616 - 0.384\,x - 0.354\,x^2 + 0.060\,x^3\ \ {\rm kHz},
  \label{e:flam}
\end{equation}
with $x=\log_{10}\Lbar$, holding to $0.41\%$ rms and $0.97\%$ at worst; the
quarkyonic models follow
\begin{equation}
  f_f M_{1.4} = 4.394 + 0.060\,x - 0.552\,x^2 + 0.085\,x^3\ \ {\rm kHz},
  \label{e:flamq}
\end{equation}
to $1.10\%$ and $2.38\%$. This reproduces, for a grid that includes quarkyonic
matter, what Sotani and Kumar~\cite{SotaniKumar21} found for nucleonic and
hybrid stars: the $f$ mode is predicted from $\Lbar$ alone to about a per cent
for the controls and to $1.1\%$ rms for the quarkyonic models. An $f$-mode frequency and a measured $\Lbar$ are therefore a consistency
test of the whole family at the per-cent level without any radius; and by the
same token the $f$ mode cannot distinguish the two families this way either.
The $g_1$ residuals are $35\%$ and $46\%$, essentially the values found
against $C$. Changing the abscissa from a structural variable to another
structural variable cannot help, because what the $g$ mode responds to is not
structural.

\begin{table}[t]
\caption{Residuals of the cubic relations of this section against the
compactness and against $\log_{10}\Lbar$, for the identical sample
($M\ge1.0\,\Msun$, $N=123$ quarkyonic and $59$ nucleonic models), and the same
statistic $\Delta=|f/f_{\rm fit}-1|$. All frequencies are scaled by
$M_{1.4}=M/1.4\,\Msun$.}
\label{t:lam}
\begin{ruledtabular}
\begin{tabular}{llcccc}
 & & \multicolumn{2}{c}{versus $C$} & \multicolumn{2}{c}{versus $\log_{10}\Lbar$} \\
Mode & Family & rms & max & rms & max \\
\hline
$f$     & nucleonic  & $0.40\%$ & $1.1\%$ & $0.41\%$ & $0.97\%$ \\
        & quarkyonic & $0.43\%$ & $0.82\%$ & $1.10\%$ & $2.4\%$ \\
$p_1$   & nucleonic  & $1.3\%$ & $3.0\%$ & $2.0\%$ & $5.0\%$ \\
        & quarkyonic & $6.8\%$ & $19\%$ & $8.1\%$ & $21\%$ \\
$g_1$   & nucleonic  & $35\%$ & $75\%$ & $35\%$ & $75\%$ \\
        & quarkyonic & $47\%$ & $117\%$ & $46\%$ & $111\%$ \\
\end{tabular}
\end{ruledtabular}
\end{table}

One caution about that comparison: quote the frequency, not the scaled
frequency. A quarkyonic star of a given $\Lbar$ is heavier than the nucleonic
control built on the same interaction (at $\Lbar=400$ the pairs are $1.44$ against $1.13\,\Msun$ for $L=30\mev$, $1.50$ against $1.35\,\Msun$ for $L=50\mev$ and $1.55$ against $1.46\,\Msun$ for $L=70\mev$), so any positive
power of the mass in the scaling works against the frequency suppression and
hides part of it. For $L=50\mev$ at $\Lbar=400$ the raw $f_{g_1}$ ratio is
$0.85$ while the ratio of $f_{g_1}M_{1.4}$ is $0.94$. Taking the raw
frequencies, at $\Lbar=400$ the quarkyonic $g_1$ sits at $0.34$, $0.85$ and
$0.78$ of its control for $L=30$, $50$ and $70\mev$, and at $\Lbar=200$ at
$0.32$, $0.80$ and $0.70$; over the same comparison the $f$ mode moves by $18\%$,
$7\%$ and $4\%$ and the $p_1$ mode by $4\%$, $3\%$ and $1\%$. The $g$ mode is
again the only one that moves by a factor.

One null result in that comparison is worth stating explicitly, because it
bounds what the test can do. The $n_t=0.5\fm$ models are indistinguishable from
their controls at every $\Lbar\ge200$ we can compare; all three ratios equal
unity to three digits, and the matched masses agree to $0.01\,\Msun$. The
reason is that a star with $\Lbar=200$ has a central density of only
$0.51$--$0.54\fm$, barely above $n_t$, so its quarkyonic core is a sliver and
the star is dynamically its own control. A tidal-deformability comparison
probes the transition only if the transition has actually occurred in the stars
being compared, and for a late onset that requires $\Lbar$ well below the range
GW170817 populated.

\subsection{The mass derivative}
\label{s:obs}

Section~\ref{s:decomp} showed that the mass derivative is where the composition
signal lives. Here we ask what an observer would actually have to measure to
see it, mode by mode and at specific masses, rather than as a chord average.

\begin{figure*}[t]
\centerline{\includegraphics[width=\textwidth]{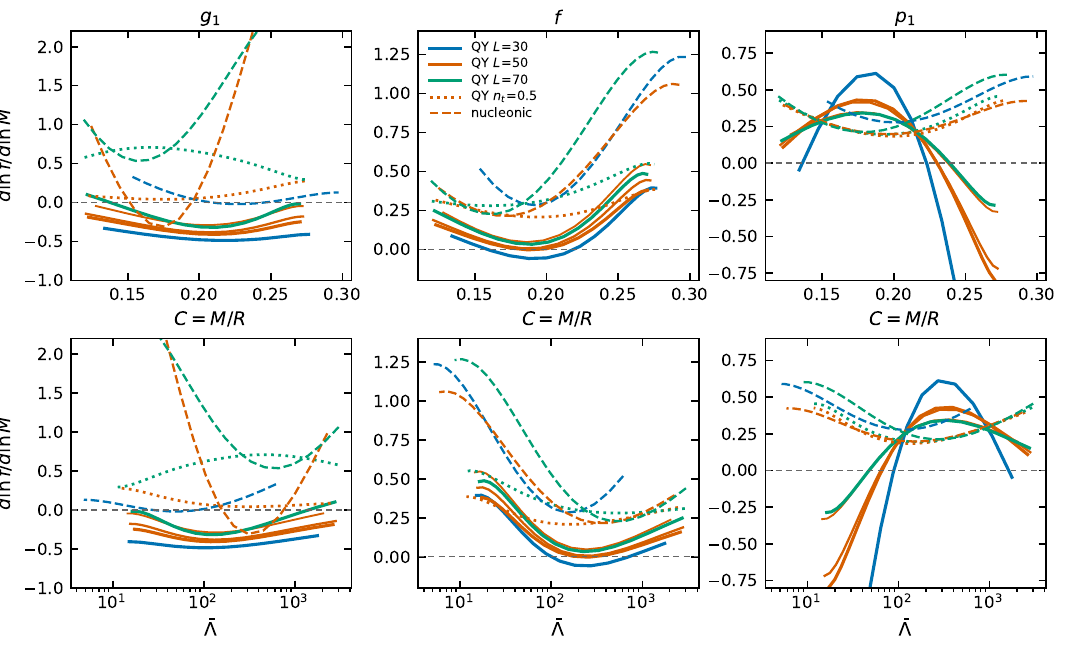}}
\caption{Logarithmic mass derivative $d\ln f/d\ln M$ of the three modes along
each stable branch, against compactness (top row), and tidal deformability
(bottom row). Solid lines are the $n_t=0.3\fm$ quarkyonic models (line width
increasing with $\Lambda$), dotted the $n_t=0.5\fm$ models, dashed the
nucleonic controls; colour encodes $L$. The derivative is taken from a cubic in
$M$ fitted to the whole branch. The dimensionless form is plotted because the
three modes differ by an order of magnitude in $\mathrm{Hz}/\Msun$.}
\label{f:dfdm}
\end{figure*}

\emph{The mass derivative separates the families where the frequency does
not.} Figure~\ref{f:dfdm} gives $d\ln f/d\ln M$ for all three modes. The
derivative is obtained from a cubic in $M$ fitted to the whole stable branch.
That fit reproduces the computed frequencies to better than $1\%$ of the mean
for every model and mode except four quarkyonic $p_1$ branches, which reach
$1.1$--$1.8\%$, and the $g_1$ branches of the $L=50$ and $70\mev$ controls
($4.3\%$ and $1.9\%$), which curve too sharply near $2\,\Msun$ for a cubic; those are also the two branches whose derivative is
largest, so the separation reported below is not sensitive to it. Three
statements follow, one per mode, and they are independent of any absolute
calibration: a systematic error common to a sequence, the Cowling shift
included, largely cancels in a logarithmic derivative.

For the $g_1$ mode the sign is the discriminator, but not at $1.4\,\Msun$: there
the controls themselves scatter across zero ($-0.30$ for $L=50\mev$, $-0.01$
for $L=30\mev$, $+0.54$ for $L=70\mev$), because the crust and the shell
compete at low mass. By $1.8\,\Msun$ the separation is clean. All five
$n_t=0.3\fm$ quarkyonic models have $d\ln f_{g_1}/d\ln M$ between $-0.28$ and
$-0.49$; every other model is positive, $+0.12$ and $+0.60$ for the two
$n_t=0.5\fm$ sets and $+1.6$ and $+2.0$ for the controls that reach that mass.
In frequency ratios, $f_{g_1}(1.8\,\Msun)/f_{g_1}(1.4\,\Msun)$ is $0.89$--$0.95$
for the five early-onset models and $1.11$--$1.26$ for the controls: a
separation of between $17$ and $42\%$ depending on the pair, which is the
precision a two-mass measurement would
have to reach.

For the $f$ mode the \emph{value} is blind to the transition, as
Eq.~(\ref{e:flam}), and its quarkyonic counterpart show, but the \emph{slope} is
not, although, as Table~\ref{t:decomp} makes clear, what that slope reports
is the mass--radius relation and not the composition. Across $1.2$--$2.0\,\Msun$ the quarkyonic sets stay within
$|d\ln f_f/d\ln M|\le0.47$, and the five early-onset ones within $0.28$,
whereas the controls climb to $+1.06$ ($L=50\mev$), and $+1.19$ ($L=70\mev$) at
$2.0\,\Msun$. Equivalently $f_f(2.0)/f_f(1.4)$ is $0.99$--$1.04$ for the
early-onset models against $1.19$ for both controls. The $f$ mode of a
quarkyonic star is nearly mass-independent above $1.4\,\Msun$, and that
flatness is a signature the absolute frequency does not carry.

For the $p_1$ mode the signature is a turnover. Every $n_t=0.3\fm$ model has a
maximum in $f_{p_1}$ at a mass below the maximum-mass configuration ($2.00$, $2.08$, $2.14$, $2.16$ and $2.24\,\Msun$ for the five sets), while neither
the $n_t=0.5\fm$ models nor any control turns over anywhere on its stable
branch. Table~\ref{t:decomp} shows that this is a genuine near-cancellation between
the two terms of Eq.~(\ref{e:decomp}): for the early-onset models $\Scomp$
runs from $+0.08$ to $-0.17$ and is small, but so is $\Sst=+0.30$ to $+0.45$, and the
turnover occurs where the second stops beating the first. The controls, whose
$\Sst$ is twice as large, never reach that point on a stable branch. The
turnover masses lie within reach of the observed population.

Finally, and negatively: no single variable organises these derivatives. Taking
the seven quarkyonic sets together and ranking $df/dM$ by Spearman coefficient
against seven candidate abscissae ($C$, $M$, the central baryon density, $\Lambda$, and the central values of $\ce^2$, $\cs^2$ and $\ce^{-2}-\cs^{-2}$), the strongest are $+0.58$ for $g_1$ against the central buoyancy factor,
$+0.60$ for the $f$ mode against the central density, and $-0.71$ for $p_1$
against the mass itself. Only the first is a composition variable, and none of
the three is tight enough to support a fit. That
is the same conclusion the $g_1$ relation reached, in a form that also applies
to the other two modes: the mass derivative is a composition diagnostic, and
composition is not a function of any one structural parameter.

\subsection{Radiation, damping and detectability}

\begin{figure*}[t]
\centerline{\includegraphics[width=0.86\textwidth]{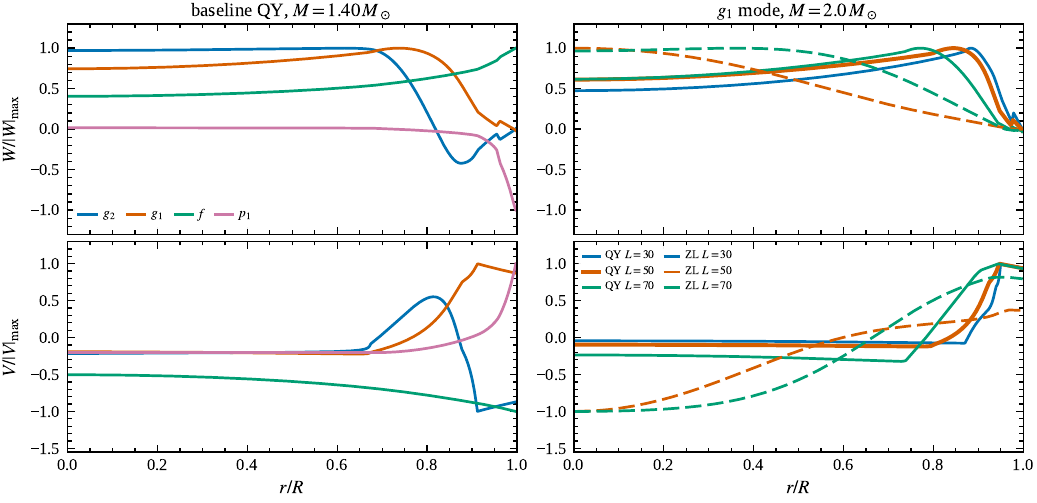}}
\caption{Radial ($W$, top), and tangential ($V$, bottom) perturbation
functions, each normalised to unit maximum modulus. Left: the $g_2$, $g_1$,
$f$ and $p_1$ modes of the baseline quarkyonic star at $1.4\,\Msun$. Right:
the $g_1$ mode of $2.0\,\Msun$ stars for several equations of state.}
\label{f:eig}
\end{figure*}

\begin{figure*}[t]
\centerline{\includegraphics[width=0.86\textwidth]{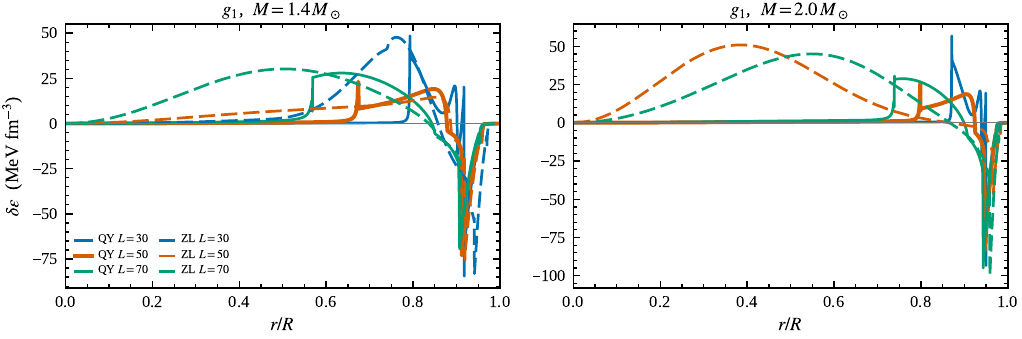}}
\caption{Eulerian energy-density perturbation of the $g_1$ mode,
Eq.~(\ref{e:depsalg}), at $1.4\,\Msun$ (left), and $2.0\,\Msun$ (right),
normalised to a mode energy of $10^{51}$~erg. In the quarkyonic stars
$\delta\varepsilon$ is driven to zero throughout the quarkyonic core and turns
on abruptly at the transition radius. The narrow oscillations at
$r/R\gtrsim0.93$ are an artefact of the piecewise-polytropic crust join.}
\label{f:deps}
\end{figure*}

Figure~\ref{f:eig} shows how the mode arranges itself around that core, and
the arrangement is not the obvious one. It would be natural to expect a
weakly stratified core simply to be left out of the motion. It is not: the
radial function $W$ is \emph{large} there, running flat between $0.75$ and
$0.95$ of its maximum from the centre out to the transition radius, with no
nodes. What the core does not do is deform. The tangential function $V$ sits
at a fifth of its maximum and barely varies across the whole core
($-0.19$ to $-0.23$ in units of its peak), changes sign at $r/R\simeq0.75$,
and only reaches full amplitude in the outer shell.

Physically the quarkyonic core is displaced bodily, almost as a rigid plug,
while the horizontal flow that a $g$ mode lives on is confined to the shell
outside it. That is what a region of vanishing $N^2$ must do: with no
buoyancy there is no restoring force to convert radial displacement into the
transverse circulation that carries the mode, so the core rides along instead
of oscillating. The hadronic controls, whose $N$ is finite everywhere, spread
both functions over the outer half of the star with no such separation.

The consequence for the energetics is quantitative. At $1.4\,\Msun$ the
quarkyonic core occupies $31\%$ of the stellar volume but holds only $14.5\%$
of the mode energy $\int(dE/dr)\,dr$; at $2.0\,\Msun$ it occupies $51\%$ and
still holds only $15.0\%$. The mode is not merely suppressed; it is
geometrically expelled, and the heavier the star the more of it is expelled
from.

Figure~\ref{f:deps} carries this into the radiation. Because the second term
of Eq.~(\ref{e:depsalg}) is proportional to $\ce^{-2}-\cs^{-2}$, which is what
collapses at $n_t$, and the first is proportional to the Eulerian $\delta p$,
which the rigid-plug motion of the core keeps small, $\delta\varepsilon$ falls
three orders of magnitude below its shell value inside the core: at $r=0.1R$
it is $3.5\times10^{-4}$ of its maximum in the shell, rising to $6\%$ of that
maximum at the transition radius, where it switches on as a step. Weighted by the $r^4$ of the
quadrupole integrand, which favours precisely the outer regions, the core
contributes $1.8\%$ of $\int|r^4\delta\varepsilon|\,dr$ at $1.4\,\Msun$ and
$4.0\%$ at $2.0\,\Msun$. A quarkyonic core is not just non-buoyant; it is
very nearly invisible to gravitational-wave emission, and that is what
lengthens the damping times below.

\begin{figure*}[t]
\centerline{\includegraphics[width=0.86\textwidth]{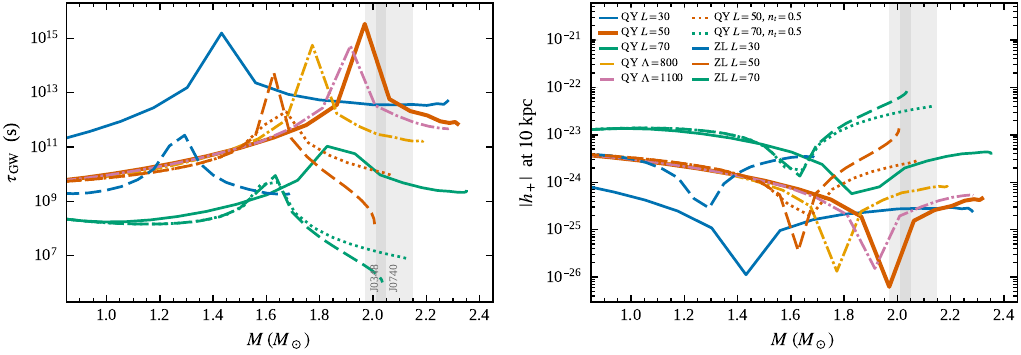}}
\caption{Gravitational-wave damping time (left), and strain amplitude at
$10$~kpc for a mode energy of $10^{51}$~erg (right) of the $g_1$ mode, against
stellar mass; curves as in Fig.~\ref{f:eos}, vertical bands as in
Fig.~\ref{f:fm}. Both quantities pass through sharp excursions at the masses
where the sign-alternating integral $\int r^4\delta\varepsilon\,dr$ cancels,
and should be read on their smooth envelope rather than point by point.}
\label{f:tau}
\end{figure*}

Figure~\ref{f:tau} shows $\taugw$ and $|h_+|$.

The damping times are long, $10^{6}$--$10^{15}$~s (absolute values carrying the factor-of-two systematic calibrated in Sec.~\ref{s:num}, which is irrelevant on this logarithmic scale), so GW emission is not the
mode's dominant damping channel; weak-interaction relaxation and viscosity will
act sooner~\cite{Reisenegger92}. That is a statement about the mode's lifetime,
not about its detectability, which is governed by the strain. For scale, Zhao
and Lattimer~\cite{ZhaoLat22} find $\taugw>10^{4}$~s for the discontinuity $g$
mode of a hybrid star against $0.1$--$1$~s for the $f$ mode; our hadronic
controls sit two to ten orders of magnitude above their $g$-mode figure
because a continuous composition gradient is a far weaker radiator than a
density jump, and the quarkyonic models sit higher still.

The curves are not smooth. $\taugw$ spikes and $|h_+|$ passes through a deep
minimum at particular masses, in every model. This is a cancellation in the radiation
integral of Eq.~(\ref{e:pgw}), and it is physical rather than numerical:
$\delta\varepsilon$ changes sign in the outer core
(Fig.~\ref{f:deps}), the positive inner and negative outer contributions to
$\int r^4\delta\varepsilon\,dr$ are of comparable size, and at one mass they
cancel. Reference~\cite{Zheng23} reports the same behaviour and the same
caution: the GW power near such a mass is highly model dependent. Our algebraic
evaluation of $\delta\varepsilon$ removes the numerical part of that difficulty (at the $1.4\,\Msun$ baseline the integral changes by $3\times10^{-5}$ in relative terms, and $\taugw$ by $5\times10^{-4}$, when the radial resolution is doubled), but not the physical sensitivity, which is what governs the
behaviour near a cancellation mass.

Away from the cancellations the quarkyonic stars radiate systematically less,
and the comparison has to be made on the smooth envelope rather than at a
single mass. Between $1.8\,\Msun$ and the end of each branch the baseline quarkyonic model
carries $\taugw$ in the range $10^{12}$--$10^{15}$~s against
$10^{8}$--$10^{10}$~s for the hadronic control, and $|h_+|$ smaller by one to
two orders of magnitude. At $2.0\,\Msun$ the tabulated values are
$3.4\times10^{13}$~s and $6.3\times10^{-26}$ against $2.7\times10^{8}$~s and
$1.1\times10^{-23}$; the ratio at that particular mass, five orders of
magnitude, sits near the top of the excursion and should not be read as a
typical figure. At $1.4\,\Msun$, where neither model is near a cancellation,
the two damping times differ by only $50\%$. The suppression has the same
origin as the frequency shift: a quarkyonic core is both non-buoyant and
non-radiating, and increasing the mass increases its share of the star.

\begin{table}[t]
\caption{Minimum GW energy $E_{\rm GW}$ (erg), that the $g_1$ mode must radiate
to be detected at $S/N=8$, from Eq.~(\ref{e:egw}), for
$S_n^{1/2}=2\times10^{-23}\,\mathrm{s}^{1/2}$ (Advanced LIGO--Virgo), and
$10^{-24}\,\mathrm{s}^{1/2}$ (Einstein Telescope).}
\label{t:egw}
\begin{ruledtabular}
\begin{tabular}{lc|cc|cc}
 & & \multicolumn{2}{c|}{aLIGO--Virgo} & \multicolumn{2}{c}{Einstein} \\
Model & $M/\Msun$ & 10 kpc & 15 Mpc & 10 kpc & 15 Mpc \\
\hline
QY $L{=}30$ & 1.4 & $1.6\times10^{45}$ & $3.7\times10^{51}$ & $4.1\times10^{42}$ & $9.2\times10^{48}$ \\
            & 2.0 & $1.2\times10^{45}$ & $2.6\times10^{51}$ & $2.9\times10^{42}$ & $6.6\times10^{48}$ \\
QY $L{=}50$ & 1.4 & $3.1\times10^{45}$ & $6.9\times10^{51}$ & $7.7\times10^{42}$ & $1.7\times10^{49}$ \\
            & 2.0 & $2.4\times10^{45}$ & $5.3\times10^{51}$ & $5.9\times10^{42}$ & $1.3\times10^{49}$ \\
QY $L{=}70$ & 1.4 & $1.0\times10^{46}$ & $2.3\times10^{52}$ & $2.5\times10^{43}$ & $5.7\times10^{49}$ \\
            & 2.0 & $8.5\times10^{45}$ & $1.9\times10^{52}$ & $2.1\times10^{43}$ & $4.8\times10^{49}$ \\
\hline
HAD $L{=}50$ & 1.4 & $4.1\times10^{45}$ & $9.2\times10^{51}$ & $1.0\times10^{43}$ & $2.3\times10^{49}$ \\
             & 2.0 & $1.0\times10^{46}$ & $2.3\times10^{52}$ & $2.5\times10^{43}$ & $5.7\times10^{49}$ \\
HAD $L{=}70$ & 1.4 & $1.5\times10^{46}$ & $3.4\times10^{52}$ & $3.8\times10^{43}$ & $8.5\times10^{49}$ \\
             & 2.0 & $3.6\times10^{46}$ & $8.0\times10^{52}$ & $8.9\times10^{43}$ & $2.0\times10^{50}$ \\
\end{tabular}
\end{ruledtabular}
\end{table}

Table~\ref{t:egw} gives the minimum detectable mode energy. For a Galactic
source at $10$~kpc the threshold is $10^{45}$--$4\times10^{46}$~erg for Advanced
LIGO--Virgo and $10^{42}$--$10^{44}$~erg for the Einstein Telescope, in both
cases far below the $\sim10^{51}$~erg that a $g$ mode excited in a post-merger
remnant or a phase-transition-driven event might plausibly
carry~\cite{Lugones21}. A source in the Virgo cluster at $15$~Mpc needs
$3\times10^{51}$--$8\times10^{52}$~erg for Advanced LIGO--Virgo and still
$10^{49}$--$2\times10^{50}$~erg for the Einstein Telescope, so only Galactic
and Local Group events are realistic targets. Because $Q=\pi f\taugw\gg1$
throughout, Eq.~(\ref{e:egw}) reduces to $E_{\rm GW}\propto f^2$ and the
thresholds simply track the frequencies; the quarkyonic thresholds are
therefore \emph{lower} than the hadronic ones at the same mass and the same
$L$, by a factor of about four at $2.0\,\Msun$. This is not a detection advantage, since the
quarkyonic mode also radiates far less; the discriminating observable is the
pair $(f_{g_1},M)$, and above all its slope.

A second, independent way the same frequency enters an observation is through
the dynamical tide. By Eq.~(\ref{e:res}), the $g_1$ mode of a star in a binary
is resonantly driven when the GW frequency passes through $f_{g_1}$, and the
right-hand axis of Fig.~\ref{f:fm} converts that frequency into the time
remaining before coalescence for an equal-mass $1.4+1.4\,\Msun$ system. Our
quarkyonic models resonate at $81$--$254$~Hz for $n_t=0.3\fm$, that is
$3.8$~s to $0.18$~s before merger and some $490$ to $70$ GW cycles from the end
of the inspiral; the hadronic controls resonate at $158$--$522$~Hz above $1.0\,\Msun$
(down to $130$~Hz if sub-solar-mass configurations are included), as little
as $0.026$~s and $22$ cycles out. Since the accumulated phase shift grows with
the number of cycles remaining after the resonance, a lower resonance frequency
is an advantage, and the fact that the quarkyonic resonance moves \emph{down}
with increasing mass while the hadronic one moves up is directly relevant to
searches for $g$-mode dynamical tides, whose detectability has recently been
argued to be within reach of current detectors for moderately eccentric
binaries~\cite{Takatsy26}. We stop short of quoting a phase shift, for the reason set out in
Sec.~\ref{s:noQ}, but the sign of the effect is not in doubt, and the scale the
literature uses is worth recalling. Yu and Weinberg~\cite{YuWeinberg17}
obtain $\Delta\Psi\sim10^{-3}$--$10^{-2}$~rad for the composition $g$ modes of
an ordinary neutron star, too small to measure in a single event; Jaikumar
\emph{et al.}~\cite{Jaikumar21} and Miao \emph{et al.}~\cite{Miao24} obtain
$0.1$--$1$~rad for the discontinuity mode at a strong first-order transition,
which is marginal but not hopeless. Our quarkyonic $g_1$ has a smaller
$\delta\varepsilon$, a smaller strain and a longer damping time than the
hadronic control at every mass, so its overlap with the tidal field cannot
exceed the control's: the quarkyonic phase shift belongs at or below the lower
end of that range, not the upper. A quarkyonic core makes the dynamical tide
\emph{harder} to see, and the observable that distinguishes it is the mode
frequency and its slope, not the phase it imprints.

\section{Discussion}
\label{s:disc}

\subsection{Quarkyonic crossover versus a Gibbs mixed phase}

The literature that treats $g$ modes as quark-matter probes is built almost
entirely on hybrid stars with a genuine phase transition. There, $\ce$ drops discontinuously at the
onset of the mixed phase while $\cs$ stays continuous, the buoyancy factor
spikes, and $f_{g_1}$ jumps upward by a factor of two or more. How far upward depends on
the model: the Brueckner--Hartree--Fock plus Dyson--Schwinger calculations of
Ref.~\cite{Zheng23} reach $0.6$--$0.9$~kHz, while Ref.~\cite{Jaikumar21}
reports $0.2$--$0.6$~kHz for its hybrid stars, a range that overlaps the
hadronic one at its lower end; Refs.~\cite{Constantinou21,ZhaoQNM22,Kumar23}
fall between. What the calculations agree on is the direction and the factor of
two, not the absolute frequency. Our quarkyonic stars do
the opposite: $\ce$ rises rather than falls, and the buoyancy factor collapses
rather than spiking.

One point of contact with that literature does not carry over, and we flag it
rather than leave it implicit. Zhao \emph{et al.}~\cite{ZhaoQNM22} report a linear correlation between the
$g$-mode frequency and the central lepton fraction for nucleonic stars, and
recover an approximate version of it for hybrid stars once the quark fraction
is added to the lepton one. On our grid even the nucleonic version is loose: a
linear fit of $f_{g_1}$ against the central $Y_L$ leaves an rms residual of
$29\%$ for the nucleonic controls, and $18\%$ for the early-onset quarkyonic
sets. The
likely reason is that we scan the symmetry-energy slope over
$L=30$--$70\mev$, which changes the composition \emph{gradient} at fixed
central $Y_L$, and it is the gradient and not the value that Eq.~(\ref{e:quad})
responds to. We report this as a difference to be resolved rather than a
disagreement, since their calculation is in full general relativity and ours is
not.

The contrast is sharp enough to be turned into a measurement strategy, and the
three cases are cleanly separated by the \emph{sign} of $df_{g_1}/dM$: rising
for a purely hadronic star, rising with a discontinuous jump at the mixed-phase
onset mass for a Gibbs hybrid, falling for a quarkyonic star with an early
enough onset. The absolute frequency is a poor discriminator on its own; our
quarkyonic $f_{g_1}$ spans $81$--$406$~Hz depending mostly on $L$, and overlaps
the hadronic range. What that strategy actually requires, and what it does not,
is the subject of the next subsection.

\subsection{What to measure}
\label{s:strategy}

Section~\ref{s:decomp} changes the observational question, and it is worth
being explicit about how, because the change cuts both ways.

The bad news first: the sign of $df_{g_1}/dM$ is not a clean composition
diagnostic. By Eq.~(\ref{e:decomp}), that sign is the sum of a compositional
term and a structural one, and the structural term is large and positive for
every star. A falling $f_{g_1}$ requires the composition term to beat it, which
on this grid needs both halves of the quarkyonic effect at once: the buoyancy
suppression, and the stiffening that stops the star from contracting. The
$n_t=0.5\fm$ models have $\Scomp=-0.57$, between three-quarters and
ninety-five per cent of the early-onset values, and still show a rising
$f_{g_1}$ because their radii shrink. A rising
$g$-mode frequency therefore does not exclude quarkyonic matter; it excludes
only the combination of an early onset and a stiff transition. A falling one is
not produced by any other case we have examined, so the test is one-sided.

The clean observable does exist, but it is not the slope. The
composition term is the slope of $f_{g_1}/\fdyn$, and $\fdyn$ is built from $M$
and $R$, both of which are measured: by radio timing and NICER for a
Galactic source, or by the inspiral phase through $\Lbar$ for a merger. An
observer who has $M$, $R$ and one $g$-mode frequency has $k=f_{g_1}/\fdyn$
directly, with no fit and no equation-of-state assumption. For a nucleonic star
$k$ is constant along the mass sequence to $2$--$9\%$
(Eq.~\ref{e:kfac}); for an early-onset quarkyonic star it falls by
$26$--$32\%$ between $1.2$ and $2.0\,\Msun$. Two sources with different masses
and measured radii therefore test the composition directly, without the
structural term ever entering. This is a stronger test than the sign of the
slope and it needs the same data.

A caution about the other two modes. It is tempting to read the
flatness of the quarkyonic $f$ mode and the turnover of $p_1$
(Sec.~\ref{s:obs}) as two further independent signatures. They are not.
Table~\ref{t:decomp} shows that their compositional terms are of the same size
and sign in both families (the matched-$L$ pairs differ by at most $59\%$, against the factor of several that separates the $g_1$ entries), and that the
difference in their observable slopes is inherited almost entirely from
$d\ln R/d\ln M$. Anything they can
tell us about the interior, a mass--radius measurement tells us more directly
and with instruments that already exist. Their value is as a consistency check
on the structure, and as the part of the signal that would survive into a
merger remnant, where buoyancy does not.

Finally, one prediction here needs no asteroseismology at all. The structural term
is itself the sharpest quarkyonic signature we have found, and it is
independent of everything else in this paper. Our early-onset sequences have
$d\ln R/d\ln M=+0.035$ to $+0.133$: the star \emph{grows} as mass is added,
by $1.8$ to $7.0\%$ between $1.2$ and $2.0\,\Msun$, where the nucleonic
controls shrink by $12$--$14\%$. A radius that does not fall between a
$1.4\,\Msun$ and a $2.0\,\Msun$ source is not by itself proof of quarkyonic
matter, any sufficiently stiff high-density behaviour will do it, but the
combination of a flat or rising $R(M)$ with a falling $f_{g_1}$ is, on this
grid, produced by nothing else.

\subsection{Fast equilibration suppresses, it does not forbid}
\label{s:fast}

There is an objection to answer. Tonetto and Lugones~\cite{Tonetto20} showed
that at a sharp interface, rapid conversion between the two phases drives the
discontinuity $g$ mode to zero frequency: the displaced element re-equilibrates
across the interface and there is nothing left to restore it. The opposite,
slow-conversion limit retains the mode and is the one usually
adopted~\cite{RaneaSandoval18,Rodriguez26}. Since we assume
the quark--nucleon equilibration is fast, why does the same argument not remove
our mode entirely?

Because our transition is continuous and the fast equilibrium is partial. Rapid
conversion at a sharp interface removes the \emph{whole} density contrast that
provides the buoyancy, so the mode disappears. In quarkyonic matter the strong
conditions remove only the quark contribution to the compositional freedom; the
lepton fractions still relax through weak processes and still stratify the
star. What survives is a $g$ mode supported by $\mut_e$ and $\mut_\mu$ gradients
alone, and Eq.~(\ref{e:quad}) is exactly the statement that this residual
buoyancy is positive-definite and small. Our order-of-magnitude suppression is
therefore the continuous-crossover version of their zero-frequency limit, not a
contradiction of it. The same logic explains why our result agrees with the
conclusion of Constantinou \emph{et al.}~\cite{Constantinou21} that a
\emph{detected} $g$ mode would point to a first-order transition rather than a
crossover: our crossover $g$ mode is harder to detect than the hadronic one,
not easier.

\subsection{Why we do not quote a tidal phase shift}
\label{s:noQ}

The natural next step from a resonance frequency is a cumulative phase shift,
and we have the machinery for it: tidal overlap integrals $Q_\alpha$ built from
the same eigenfunctions, which reproduce the closed-form incompressible result
and the mode-sum Love-number rule to one part in $10^{10}$ in the Newtonian
incompressible limit where both are known analytically. On the relativistic
stars themselves that sum rule is not recovered: the $f$ mode supplies only
$39$--$61\%$ of it, where in the analytic limit it supplies almost all. We do not
quote a number, because for the $g$ mode the integral is a near-cancellation
and we cannot control it. Two diagnostics say so. The fraction of $Q_{g_1}$
generated inside the transition radius comes out at $1.6$--$1.8$: a fraction
larger than one, which can only mean that the inner and outer contributions
have opposite signs and nearly cancel. And the mode-sum rule, which the $f$
mode should dominate, is not recovered at all on these stars. Both are what a
difference of large and nearly equal contributions looks like, and our
barotropic crust and our radial resolution are not good enough to resolve it.
Quoting a $g$-mode overlap from this machinery would be reporting a
cancellation error. What can be said without the integral is said in
Sec.~\ref{s:res}: the quarkyonic mode radiates less and couples less than the
hadronic one at every mass, so its phase shift lies at or below the
$10^{-3}$--$10^{-2}$~rad that Ref.~\cite{YuWeinberg17} obtains for an ordinary
neutron star.

\subsection{Limitations}
\label{s:lim}

The results hold on the grid explored here and within the approximations
listed below. For each we give the direction of the error, its size where we
can estimate it, and whether the central claim, the sign of $\Scomp$ for the $g_1$ mode, survives it.

\emph{Cowling.} We neglect metric perturbations. For $g$ modes the error
reaches $10\%$ and grows with mass, always in the sense that Cowling
\emph{underestimates} the frequency~\cite{ZhaoQNM22,SotaniTogashi26}; for the
$f$ and $p$ modes it goes the other way and reaches
$30\%$~\cite{Yoshida97,Chirenti15}, though on our own stars it is
$5.8$--$16.2\%$ (Sec.~\ref{s:num}). The
numerical coefficients of Eq.~(\ref{e:ffit}) should therefore not be compared
with full-GR fits, although the $0.61\%$ scatter about it (a statement about the spread among equations of state at fixed compactness, not about the absolute frequency) should be far less affected, since the Cowling error is
common to the two families. We have not verified that cancellation in full
general relativity, and the internal comparisons of Sec.~\ref{s:univ} inherit
the same caveat.

The decomposition of Sec.~\ref{s:decomp} localises this uncertainty usefully.
The structural term $\Sst$ is built from $M$ and $R$ alone, which come from
the Tolman--Oppenheimer--Volkoff equations and are exact: the Cowling
approximation cannot touch it. The entire error is confined to $\Scomp$.
Applying the correction shape used in Sec.~\ref{s:sum} (zero at $1.2\,\Msun$ rising to $10\%$ at the maximum mass) adds
$\ln(1.10)/\ln(2.2/1.2)\simeq+0.16$ to every $\Scomp$ in
Table~\ref{t:decomp}, the same shift for both families. The early-onset
quarkyonic values move from $-0.60$--$-0.75$ to $-0.44$--$-0.59$ and the
controls from $-0.04$--$+0.16$ to $+0.12$--$+0.32$. The separation, and the
sign, survive a correction of that size; what would not survive is a Cowling
error whose \emph{mass dependence} differed between the two families by more
than about $0.5$ in logarithmic slope, and we have no reason to expect one but
no proof against it either.

\emph{Estimators and diagnostics.} Two distinct estimators of the mass
derivative are used and they should not be confused. Table~\ref{t:decomp} reports chord averages between $1.2$ and
$2.0\,\Msun$, which involve no fitting and are exact up to the interpolation of
$f$ onto those two masses; the decomposition itself is an identity and closes
to machine precision. Figure~\ref{f:dfdm} and the
fixed-mass slopes quoted with it come from a cubic in $M$ fitted to the whole
branch, which reproduces the computed frequencies to better than $1\%$ of the
mean except on four quarkyonic $p_1$ branches ($1.1$--$1.8\%$), and the two
hadronic $g_1$ branches ($4.3\%$ and $1.9\%$). Local
slopes near $1.4\,\Msun$ are the least reliable numbers in the paper, which is
why the discriminator is quoted at $1.8\,\Msun$ and as a chord average.

The WKB cavity integral at the end of
Sec.~\ref{s:decomp} is likewise a consistency check and not a derivation. The
fundamental $g$ mode is not in the asymptotic regime, the ratio
$f_{g_1}/\mathcal{I}_N$ ranges from $1.0$ to $2.5$ across the grid, and the
quarkyonic fall it predicts ($3\%$ for the baseline) is a quarter of the fall
in the eigenfrequency. We use it only to show that the sign reversal is already
present in the cavity integral, not to predict frequencies.

\emph{Damping channels.} We compute only GW damping. Weak-interaction
relaxation and shear viscosity are expected to dominate at these frequencies,
and we make no claim about the mode's actual lifetime. The quadrupole estimate
of the radiated power itself is low by $8$ to $40\%$ where we can calibrate it
in Sec.~\ref{s:num}, against the Andersson--Kokkotas $f$-mode fit, and carries
no tight internal check of its own; the ratios between models, which are what
we quote, are unaffected.

\emph{The crust.} We set $N=0$ in the crust and splice SLy at $\nB=0.08\fm$,
leaving a $\sim1\%$ mismatch in $\varepsilon$. The join supports a spurious
interface mode that can lie close to $g_3$ in some models, produces the narrow
spike in $N$ excluded from Fig.~\ref{f:internal} and the narrow features at
$r/R\gtrsim0.93$ in Fig.~\ref{f:deps}, and is one of the two reasons the tidal
overlap is unreliable. None of this affects $g_1$, $g_2$, $f$ or $p_n$. A
realistic inner crust would instead support genuine crust $g$ modes that can
cross the core modes~\cite{Sun25}; a barotropic crust removes them, which keeps
our mode classification clean at the cost of not describing that physics.

\emph{One control with a barotropic core.} In the purely hadronic ZL
interaction with $L=30\mev$ the symmetry energy turns over so sharply that the
lepton fractions reach zero at $\nB=0.78\fm$. Above that density the matter is
pure neutron matter, the quadratic form (\ref{e:quad}) vanishes identically,
and $\cs=\ce$ exactly. The positive-definiteness condition is approached but
never violated, here or anywhere else on the grid, so nothing was flagged or
excluded; we record this because a reader may expect the turnover to produce
an instability, and it does not. What it does produce is a star with a
buoyancy-free inner core: at $1.4\,\Msun$ the central density is $0.97\fm$,
so the $g$ mode of that model already lives outside a lepton-free centre. The
model has no entry above $1.6\,\Msun$ in Table~\ref{t:fix} or at
$2.0\,\Msun$ in Table~\ref{t:modes} for the simpler reason that its maximum
mass is $1.69\,\Msun$. The corresponding quarkyonic model is unaffected,
because the quarks appear well below that density.

\emph{Finite reaction rates and the high overtones.} We treat the weak
reactions as strictly frozen. They are not: they proceed at a finite rate, and
Andersson and Pnigouras~\cite{Andersson19} and Counsell \emph{et
al.}~\cite{Counsell24} have shown that this gives the mode frequencies a small
imaginary part and suppresses the high-order composition $g$ modes, which
otherwise form an infinite spectrum. Our $g_3$ is precisely the kind of mode
their argument removes, and we quote it only to establish the ordering of the
spectrum and the node counts; every physical conclusion in this paper rests on
$g_1$, whose period is short compared with the modified-Urca timescale at the
temperatures where a cold neutron star lives. For the same reason we do not
attach significance to the $g_3$ frequencies in Table~\ref{t:modes} beyond
their role as a numerical check. The competing rate is set by the weak
processes during inspiral, quantified by Arras and Weinberg~\cite{Arras19}, and
by particle diffusion~\cite{Kantor24}.

\emph{Microphysics.} Zero temperature, no superfluidity, no rotation, no
magnetic field, no hyperons, and one functional form for the nucleon
interaction. Each of these has a known effect on hadronic $g$ modes and none
has been examined for a quarkyonic core.

Superfluidity is the sharpest of them. Andersson and Comer~\cite{Andersson01}
argued that a superfluid core supports no independent propagating $g$ modes at
all. Kantor and Gusakov~\cite{Kantor14} then showed that the modes return once
muons are included, with the buoyancy carried by the $n_\mu/n_e$ gradient
rather than by the proton fraction, and frequencies reaching $\sim500$~Hz; see
also Refs.~\cite{Gusakov13,Passamonti16,Dommes16}. Our matter contains muons and our buoyancy is already leptonic, so the
quarkyonic suppression should survive superfluidity in form. What
Sec.~\ref{s:chan} adds to that expectation is a sign: the rescue channel
Kantor and Gusakov rely on is precisely the minority channel in a quarkyonic
core, so superfluidity is more likely to deepen the suppression than to lift
it. Settling the question needs entrainment and a two-fluid
treatment~\cite{RauWasserman18}, which we have not attempted.

Finite reaction rates work the other way. Counsell \emph{et
al.}~\cite{Counsell24} show that nuclear reactions drive the matter back
towards $\beta$ equilibrium during the oscillation itself, which erodes the
very stratification the mode lives on: the frequencies acquire an imaginary
part and the high overtones are overdamped out of the spectrum altogether.
Our modes are computed in the opposite, non-reacting limit, so that mechanism
can only reduce the buoyancy we find, not restore it, and it removes the
overtones first, which is why $g_3$ is used here only as a node-count check.
Temperature sets the other boundary. Composition $g$ modes are the relevant
modes for a cold star; above a few MeV the weak reactions become resonant with
the mode and bulk viscosity suppresses them
outright~\cite{Lozano22,ZhaoWarm25}. Zhao \emph{et al.}~\cite{ZhaoWarm25}
generalise the adiabatic sound speed to a complex, frequency-dependent
\emph{dynamical} sound speed for exactly this reason. Our results apply to
cold, catalysed matter; a merger remnant is a different problem, and the $f$
and $p_1$ signatures reported in Sec.~\ref{s:univ} are the ones that would
survive into it, because they do not rely on buoyancy.

\emph{Model dependence of the shell.} We use the Zhao--Lattimer per-species
momentum shell with chemical equilibrium between quarks and nucleons. Other
realisations would change $\ce^2$ and $\cs^2$ in detail: the universal shell
of Ref.~\cite{McLerran19}, the duality-based constructions of
Ref.~\cite{Fujimoto24}, or the RMF-based treatment used for $\omega$ modes in
Ref.~\cite{Dey26}. What we expect to survive is the structural statement,
since it follows from Eq.~(\ref{e:master}): as long as quark--nucleon
equilibration is a strong process, the quarks carry no independent
thermodynamic freedom on a mode period, and the buoyancy is suppressed wherever
the matter is simultaneously stiff and compositionally rigid.

\section{Summary}
\label{s:sum}

We have derived the adiabatic sound speed of $\beta$-equilibrium quarkyonic
matter from a single observation: quark--nucleon chemical equilibrium is
maintained by the strong interaction and therefore survives a $g$-mode period
intact, while the lepton fractions do not move at all. The energy variation then collapses to
$d\varepsilon=\mu_n d\nB+\mut_e dn_e+\mut_\mu dn_\mu$, with the quarks absent,
and the difference of the squared sound speeds becomes the positive-definite
quadratic form (\ref{e:quad}), which doubles as a convective-stability
diagnostic. Above the transition the nucleon momentum shell is full, so the
conventional frozen-composition prescription is singular there and the
construction we use is the only regular one. We verified the formalism against
an independent evaluation of $\cs^2-\ce^2$ to $10^{-7}$--$10^{-4}$ and the mode
solver against an independent integration of the Thorne system to $10^{-6}$ in
frequency, and computed the structural and gravitational-wave observables for
ten parameter sets.

Everything a quarkyonic core does to the oscillation spectrum follows from a
single fact: the core is only weakly stratified. The buoyancy factor collapses at
the onset by a factor of $9.5$ to $19$ for an early transition, $284$ in the
softest set, and by only $2$ to $3$ for a late one, which is why the late-onset
models keep a rising $f_{g_1}$. The cause is not that $\cs^2-\ce^2$ changes
much, which it does not, but that $\ce^2\cs^2$ jumps by one to two orders of
magnitude. The $g$ mode is then confined to the nucleonic shell outside the
core, and the shell thins as the mass grows. The frequency therefore falls, from
$158$--$522$~Hz in the hadronic controls to $81$--$254$~Hz for $n_t=0.3\fm$,
and it falls \emph{with increasing mass}, monotonically from $1.2\,\Msun$ to
within $0.03\,\Msun$ of the maximum mass, where the controls rise without
exception. A late onset, $n_t=0.5\fm$, suppresses the frequency but leaves the
slope positive: the reversal needs a core that occupies an appreciable fraction
of the star. Because $\delta\varepsilon$ is driven to zero through that core,
the mode also stops radiating, and the damping time lengthens by orders of
magnitude while the strain falls.

The other two modes are far less useful, and there is a way of saying how much
less that removes the arbitrariness of choosing which quantity to hold fixed.
Write each frequency as the star's dynamical frequency
$\fdyn=(2\pi)^{-1}(GM/R^3)^{1/2}$ times a dimensionless remainder, as in
Eq.~(\ref{e:decomp}). Measured that way the $f$ and $p_1$ modes of a quarkyonic
star are ordinary: their remainders vary with mass much as the controls' do, and the
whole of their anomalous mass trend is inherited from the mass--radius
relation, which in these models is anomalous in its own right: the stars
\emph{expand} by up to $7\%$ between $1.2$ and $2.0\,\Msun$ where the controls
contract by $12$--$14\%$. Compared with its own control at the same
compactness, the quarkyonic $f$ mode moves by less than $2\%$, in agreement
with what has been found for a quark--hadron crossover~\cite{SotaniKojo23}, and
the $p_1$ mode by up to a fifth, but only at high compactness, and at fixed
mass the two families agree on $f_{p_1}$ to about $3\%$. Whatever those two
modes can say about the interior, a mass--radius measurement says more directly.

The $g_1$ remainder is the exception, and it is the reason for this paper. It
is constant to $2$--$9\%$ along a nucleonic sequence, so that a hadronic $g$
mode is little more than $\fdyn$ times a number fixed by the symmetry energy,
and it falls by $26$--$32\%$ along an early-onset quarkyonic one; the frequency
itself moves by a factor of two at $2.0\,\Msun$. If quarkyonic matter is to be
found anywhere in the oscillation spectrum, the $g$ mode is where to look, and
that single dimensionless number (not the frequency, and not its slope) is
the cleanest thing to aim at: it needs a mass, a radius and one $g$-mode
frequency, and no equation-of-state assumption anywhere.

That conclusion must be read together with the approximation on which it
rests. Our
frequencies are relativistic Cowling values and therefore lower bounds, by up
to about ten per cent and by more at higher mass~\cite{ZhaoQNM22}. Applying a
correction of that shape (zero at $1.2\,\Msun$, ten per cent at the maximum mass) to both families leaves the net decrease intact in all five
$n_t=0.3\fm$ models, $-2\%$ to $-15\%$ instead of $-11\%$ to $-22\%$, but it
removes pointwise monotonicity in two of them. The sign of the trend survives;
its smoothness is not yet established, and settling that needs the metric
perturbations.

The practical statement is a differential one, and it is the one we would
stake the paper on. A single $g$-mode frequency
constrains little, because the absolute value is set mostly by the
symmetry-energy slope and the quarkyonic and hadronic ranges overlap. Two
frequencies at different masses, or one frequency with an independent mass,
separate three cases by the sign of $df_{g_1}/dM$: rising for a purely hadronic
star, rising with a jump at the onset mass for a Gibbs hybrid, falling for a
quarkyonic star with an early transition. The reference mass matters: at
$1.4\,\Msun$ the hadronic controls themselves scatter across zero, whereas by
$1.8\,\Msun$ every early-onset quarkyonic model has
$d\ln f_{g_1}/d\ln M$ between $-0.28$ and $-0.49$ and every other model is
positive. The same differential logic applies to the other two modes and is
worth recording, because it does not require the $g$ mode to be detected at
all: the quarkyonic $f$ mode is nearly flat in mass above $1.4\,\Msun$
($f_f(2.0)/f_f(1.4)=0.99$--$1.04$, against $1.19$ for both controls), and the
$p_1$ frequency turns over at $2.00$--$2.24\,\Msun$ in every early-onset model
while no control turns over at all. These differential tests need masses but no
radius, which is their attraction; the dimensionless remainder above is the
sharper diagnostic but needs a radius or a tidal deformability as well. Either
way the abscissa is not the obstacle: recast against $\Lbar$, which an inspiral
measures directly, the $f$-mode relation still holds to about a per cent and
the $g_1$ relation still fails by tens of per cent, exactly as against the
compactness (Table~\ref{t:lam}). Whether that measurement is within
reach is a separate question, and on present evidence the answer is not
encouraging for a single event; but the eccentric-binary channel of
Ref.~\cite{Takatsy26} and third-generation detectors change the arithmetic, and
a Galactic event would change it completely.

\section{Acknowledgments}
PJK and BK thank Tianqi Zhao for their invaluable insights, guidance and feedback, which significantly shaped this research and enabled its successful completion

\end{document}